\documentclass[journal=jctcce,manuscript=article]{achemso}
\usepackage{amssymb,amsmath,amsfonts,multicol,multirow,longtable,array,mathpazo}
\usepackage{lscape}
\usepackage[version=3]{mhchem} 
\usepackage{appendix}
\usepackage{soul} 
\usepackage[usenames]{color}
\usepackage{makecell}
\usepackage{enumerate}
\usepackage{xr}
\usepackage[T1]{fontenc} 
\usepackage{booktabs}
\usepackage{tikz}
\usetikzlibrary{shapes.geometric}
\usepackage{threeparttable}
\usepackage{graphicx}
\usepackage{subfigure}
\usepackage{colortbl}
\usepackage{framed}
\usepackage{verbatim}
\usepackage{amsbsy}
\usepackage{bm}
\usepackage{bbm}
\usepackage[normalem]{ulem}
\usepackage{float}
\usepackage{algorithm}
\usepackage{algpseudocode}
\usepackage{float}
\usepackage{subfigure}
\usepackage{simplewick} 
\usepackage{siunitx}
\usepackage{rotating}
\usepackage{xr}
\newcounter{xscheme}

\newfloat{Algorithm}{htbp}{alg}
\floatname{Algorithm}{Algorithm}

\newcounter{exe}[figure]
\newcommand{\iexe}{\refstepcounter{exe}\the\value{exe}:}

\setkeys{acs}{maxauthors = 0} 

\author{Ning Zhang}\email{ningzhang1024@gmail.com}
\affiliation{Qingdao Institute for Theoretical and Computational Sciences and Center for Optics Research and Engineering,
	Shandong University, Qingdao, Shandong 266237, China}
\author{Wenjian Liu}\email{liuwj@sdu.edu.cn}
\affiliation{Qingdao Institute for Theoretical and Computational Sciences and Center for Optics Research and Engineering,
	Shandong University, Qingdao, Shandong 266237, China}

\title{Multireference Second-Order Perturbation Theories Based on Iterative Configuration Interaction with Selection}

\setkeys{acs}{articletitle=true}

\begin{document}
	
\begin{abstract}
Accurate description of the electronic structure of strongly correlated systems of electrons necessitates a balanced treatment of
the static and dynamic components of correlation. Under the assumption that the static correlation can well be described
by selected configuration interaction, to the extent that the residual dynamic correlation can be accounted for accurately
by a second-order perturbation theory (PT2), a family of multireference PT2s (iCI-MRPT2)
based on the iterative configuration interaction with selection [SiCI; J. Chem. Theory Comput. 16, 2296 (2020)]
are proposed and compared. The first step is to SiCI-partition a large complete active space (CAS) $P$ into
the direct sum of a primary ($P_m$) and a secondary ($P_s$) subspace via a parameter $C_{\mathrm{min}}$.
The former serves as the actual reference space, whereas the latter is treated via Epstein-Nesbet PT2.
An MRPT2 is then invoked to account for the dynamic correction of the $Q=1-P$ space to the $P_m$ space.
Different MRPT2s are distinguished by the underlying zeroth-order Hamiltonians.
To circumvent the prohibitive computational cost and memory footprint associated with high-order reduced density matrices
in the fully internally contracted scheme and the exponential growth of external functions (XF) in the uncontracted scheme,
a partial contraction strategy is employed for constructing the XFs: those
involving one or two inactive (core or virtual) orbitals are kept uncontracted, whereas
those involving three or four inactive orbitals are internally contracted.
They span the uncontracted ($Q_u$) and contracted ($Q_c$) portions of $Q$, respectively.
The former is further split into two subspaces $Q_u^c$ and $Q_u^u$ via a parameter $Q_{\mathrm{min}}$,
which are to be treated in coupled and uncoupled manners, respectively.
Extrapolations to zero $Q_{\mathrm{min}}$ and $C_{\mathrm{min}}$ can be performed, so as to approach CAS-based MRPT2.
A unified and efficient implementation of the iCI-MRPT2s has been achieved via the MetaWave platform [J. Phys. Chem. A 129, 5170 (2025)].
Calculations with active spaces as large as CAS(56e,58o) can now be performed,
as showcased by the singlet-triplet gaps in polyacenes, the triplet-quintet gap in iron porphyrin, and
the potential energy curve of the chromium dimer.
The size-consistency of the iCI-MRPT2s is also analyzed in detail.

\end{abstract}

\maketitle

\clearpage
\newpage

\section{Introduction}
Just like single-reference second-order perturbation theory (SRPT2) is the first step to go beyond
the mean-field description of the electronic structure of weakly correlated systems,
multireference second-order perturbation theory (MRPT2) is the first meaningful step that can be taken
to go beyond the multiconfigurational self-consistent field (MCSCF) description of the electronic structures of strongly correlated systems.
However, the situation is much more involved in the latter. Apart from the choice of a proper zeroth-order Hamiltonian
as in SRPT2, both the reference space and the first-order interacting space (FOIS) must be designed with care.
As pointed out recently, a good MRPT2 should satisfy as many as possible of the ten criteria\cite{PASPT2New},
the very first of which is the adaptive treatment of the static and dynamic components of electron correlation, so as to account for
the strong entanglement and mutual interchangeability of the two components in cases of many active electrons and active orbitals.
This means essentially that one should give up structured and hence inflexible reference spaces.
Instead, a suitable reference space should be constructed automatically to adapt to
the varying near-degeneracy of the target states across different geometries.
The most direct means for such a purpose is obviously selected configuration interaction (sCI),
which has a long history\cite{CIPSIa} and has been revived in the last decade\cite{HBCI2016,HBCI2017a,HBCI2017b,HBCI2017c,SHBCI2018,HBCICASSCF,HBCIOrbOpt2021,4C2C-HCI2023,HBCIMPI,
ASCI2016,ASCI2017,ASCI2018PT2,ASCISCF,ASCIMPI,
ProjCI,ACI2016,ACI2017,ACI2018,
ICE-1,ICE-2,iCI,iCIPT2,iCIPT2New,iCISCF}. However, all these approaches
have been designed as an effective solver of complete-active-space/full configuration interaction (CASCI/FCI), where
the Epstein-Nesbet PT2 (ENPT2)\cite{Epstein,Nesbet} is unanimously adopted to account for the (inner-space)
dynamic correlation after the selection step for static correlation.
To make this clear, we introduce the projectors $P_m$ and $P_s$ for the variational space
determined by sCI and that left over, respectively. The direct sum of the two
spaces is equal to CASCI/FCI if the selection is performed within CASCI/FCI.
It is known that ENPT2 is not suited for the dynamic correlation
of the external space $Q$ ($=1-P=1-P_m-P_s$) beyond CASCI
(vide post). To this end, a suitable
MRPT2 remains to be formulated, with which the total correlation
energy can be calculated as $E^{\mathrm{sCI(2)}}[P_s, P_m]+E^{\mathrm{MRPT2}}_c[Q,P_m]$, with
$E^{\mathrm{sCI(2)}}[P_s,P_m]=E^{(0)}[P_m]+E^{\mathrm{ENPT2}}_c[P_s,P_m]$.

\begin{figure}[!ht]
	\includegraphics[width=0.9\textwidth]{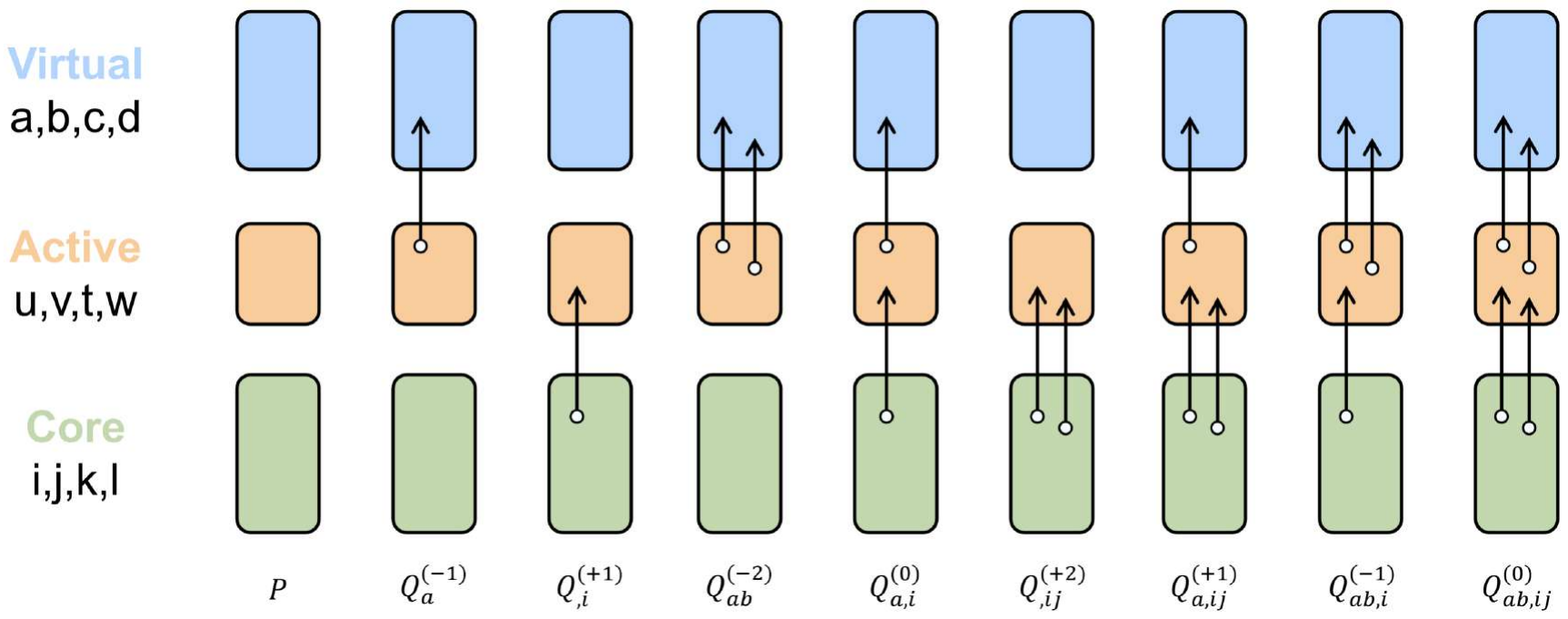}
	\caption{Illustration of external spaces $Q_l^{(k)}$ in MRPT2.}
	\label{fig:FOIS_MRPT2}
\end{figure}

Having determined the adaptive, partial active space (PAS)\cite{PASPT2} with wave function $|\Psi^{(0)}[P_m]\rangle$
and energy $E^{(0)}[P_m]$, the external functions (XFs) spanning the FOIS $Q$ ($=1-P$) remain to be defined.
Note first that the XFs can be classified into eight disjoint families characterized by the number
of electrons excited from/to the inactive (core or virtual) orbitals, see Fig.~\ref{fig:FOIS_MRPT2} and Table \ref{table:FOIS_MRPT2}.
They can be denoted symbolically as $Q_l^{(k)}$, where $l$ and $k$ refer to the occupation pattern of the inactive orbitals
and the change in the number of active electrons, respectively.
The operationally simplest XFs are obviously individual configuration state functions (CSFs)
that can be generated by single and double excitations from each reference CSF within the PAS.
However, the number $N_{\mathrm{XF}}$ of such XFs scales as $|P_m| N_{int}^2 N_v^2\sim |P_m|N^4$,
with $N_{int}=N_c+N_a$ and $N_v$ being the numbers of internal [core ($N_c$) or active ($N_a$)] and virtual (external) orbitals,
respectively, and hence becomes prohibitive for large $|P_m|$ or large basis size $N$.
To reduce this, some contractions can be introduced.
For instance, $N_{\mathrm{XF}}$ is reduced to $|P_m| N_{int}^2 N_a$ in the external contraction (where
the virtual orbitals are contracted in every excited CSF)\cite{xcMRCI2,WenImprovedEC},
to  $|P_m| N_a^3$ in the double contraction (where both the virtual and core
orbitals are contracted in every excited CSF)\cite{WenDoubleContraction},
to $N_{int}^2 N_v^2$ in the full internal contraction (FIC, where the XFs
are generated by applying distinct excitation operators $\hat{E}_M$
(cf. Table \ref{table:FOIS_MRPT2}) on $|\Psi^{(0)}[P_m]\rangle$)\cite{MeyerIC,icMRCI1b,icMRCI1c}, and even to $N_c^2 N_v^2$
in the strong internal contraction (SIC, where the active orbitals are further contracted on top of FIC)\cite{SC-NEVPT2}. It is clear that only FIC and SIC
can remove the dependence of $N_{\mathrm{XF}}$ on the potentially exponentially large $|P_m|$. However,
the price to pay here is that even the fifth-order reduced density matrices (5-RDM) have to
be handled explicitly\cite{guo2021approximations2,ICEFRNEPT2} to avoid artificial numerical instabilities\cite{cumulant2009,guo2021approximations1}
encountered in the implementation of PAS-based NEVPT2\cite{NEVPT2a}
(n-electron valence second-order perturbation theory).
Nevertheless, the computational cost can be reduced down to that required for the evaluation of 3-RDMs,
by proper factorization of all matrix elements involving the 4-RDMs and 5-RDMs\cite{NEVPT2woHighRDM}.
As a matter of fact, a RDM-free implementation of the FIC/SIC scheme has been available\cite{XianCI2018} for a long
time, where the internal contraction coefficients [$C_{q,M}=\langle \Phi_q|\hat{E}_M|\Psi^{(0)}[P_m]\rangle$
with CSFs $|\Phi_q\rangle\in Q$, which are also needed for the evaluation of 3-RDMs via resolution of the identity]
can be evaluated very efficiently
by making full use of the particle-hole correspondence in the graphical unitary group approach\cite{Wen1,Wen4,Wen6,Wen8},
which allows for pre-computing and storing the segment values of both external and core orbitals, so that the internal contraction
coefficients are reduced only to the active orbitals.
An effective pre-screening can further be introduced\cite{SDSPT2s} to
prune away a large amount (typically 25\%) of doubly excited CSFs involving three active orbitals (by the DVD approximation)
as well as batches of the CSFs with the same three active orbitals but different inactive orbitals
(by the integral-driven cutoff). The actual implementation of SDSPT2s\cite{SDSPT2s} (static-dynamic-static
second-order perturbation theory\cite{SDS,SDSPT2} with selection) does not invoke
the rank-reduction trick\cite{DyallHmat} (which holds only in the case of CAS), and
can produce the full-rank NEVPT2\cite{guo2021approximations2,ICEFRNEPT2} for free.
Note that there exist also novel implementations\cite{RR-MRCI2024,DSRG2025} of the FIC scheme,
where the high-order RDMs are converted to transition matrix elements
of lower-rank excitation operators over first- and second-order FIC residues, the number of which
can be reduced significantly by a renormalization procedure.
Recent implementations of MRPT2 (and MRCI) along these lines can handle active spaces of
around 30 electrons in 30 active orbitals \cite{ICEFRNEPT2,ASCIMRPT2,DMRG2sCI-MRCI2018,mrenpt2DMRG,DMRG2sCI-MRCI2024,RR-MRCI2024}.
However, to handle even larger active spaces, methods that require at most 2-RDMs
should be developed. Given the availability of such methods\cite{AC0,DSRG2018,DSRG2025},
more flexible and error-controllable methods are still desired. In this context,
the partial contraction framework\cite{WernerIC1988,CWcontractionJCP2000} is most appealing,
where the XFs are kept uncontracted (contracted) if the FIC of them would require higher than (at most) 2-RDMs.

\begin{table}
	\centering
	\caption{Classification of the first-order interacting space in MRPT2}
\begin{threeparttable}
\begin{tabular}{l|l|lc|cclc}\toprule
	category &$\hat{E}_M$\tnote{a}&\multicolumn{2}{c|}{subspace} & $N_p$\tnote{b} & $N_h$\tnote{b} & \multicolumn{1}{c}{$N_{\text{oCFG}}$\tnote{c}} & RDM\tnote{d} \\\toprule
	\multirow{2}{*}{internal}
	&$E_{ui}$, $e_{ui,vw}$               &$Q_{,i}^{(1)}$       & $V_A$ &0&1&$O(N_{\text{a}}^3N_{\text{c}})$             & 4\\
	&$e_{ui,vj}$                         &$Q_{,i j}^{(2)}$     & $V_B$ &0&2&$O(N_{\text{a}}^2N_{\text{c}}^2)$           & 3\\ \midrule
	\multirow{3}{*}{semi-internal}
	&$E_{au}$, $e_{au,vw}$               & $Q_a^{(-1)}$        & $V_C$ &1&0&$O(N_{\text{a}}^3N_{\text{v}})$             & 4\\
	&$E_{ai}$, $e_{ai, vu}$, $e_{vi,au}$ & $Q_{a, i}^{(0)}$    & $V_D$ &1&1&$O(N_{\text{a}}^2N_{\text{c}}N_{\text{v}})$ & 3\\
	&$e_{ai,uj}$                         & $Q_{a, i j}^{(1)}$  & $V_E$ &1&2&$O(N_{\text{a}}N_{\text{c}}^2N_{\text{v}})$ & 2\\ \midrule
	\multirow{3}{*}{external}
	&$e_{au,bv}$                         &$Q_{a b}^{(-2)}$     & $V_F$ &2&0&$O(N_{\text{a}}^2N_{\text{v}}^2)$           & 3\\
	&$e_{ai,bu}$                         &$Q_{a b, i}^{(-1)}$  & $V_G$ &2&1&$O(N_{\text{a}}N_{\text{c}}N_{\text{v}}^2)$ & 2\\
	&$e_{ai,bj}$                         &$Q_{a b, i j}^{(0)}$ & $V_H$ &2&2&$O(N_{\text{c}}^2N_{\text{v}}^2)$           & 0\\ \bottomrule
\end{tabular}
\begin{tablenotes}
	\item[a] $E_{pq}=\sum_{\sigma}a^{p\sigma}_{q\sigma}$, $e_{pq,rs}=E_{pq}E_{rs}-\delta_{qr}E_{ps}$.
    \item[b] $N_p$ and $N_h$ denote the numbers of electrons in the particle and hole spaces, respectively. $k=N_h-N_p$ in $Q_l^{(k)}$.
	\item[c] $N_{\text{oCFG}}$ indicates the scaling of
the number of orbital configurations, with $N_{\text{c}}$, $N_{\text{a}}$ and $N_{\text{v}}$ being numbers
of core, active and virtual orbitals, respectively.
	\item[d] The highest rank of reduced density matrices (RDM) required for each subspace in CAS-based FIC-MRPT2 along with rank reduction.
\end{tablenotes}
\end{threeparttable}
	\label{table:FOIS_MRPT2}
\end{table}

Motivated by the above, we here report several variants of iCI-MRPT2, which take the same
reference (determined by iterative configuration interaction (iCI) with selection\cite{iCI,iCIPT2,iCIPT2New,iCISCF})
and external spaces, but differ in the choice of zeroth-order Hamiltonians.
Apart from the partial contraction (vide post), a selection is to be performed over the uncontracted XFs.
The two together furnish high efficiency for both computation and memory footprint.

The rest of the paper is organized as follows.
Sec.~\ref{sec:theory} is devoted to iCI, iCI-MRPT2 and size-consistency of iCI-MRPT2, following the convention:
$|I\mu\rangle$ denotes
the $\mu$-th CSF of the orbital configuration (oCFG) $|I\rangle$;
core, active and virtual orbitals are denoted as $\{i, j, k, \dots\}$, $\{u, v, w, \dots\}$ and $\{a, b, c, \dots\}$,
respectively. The implementation of iCI-MRPT2 is described in Sec. \ref{sec:impl}.
The potential energy curve of the chromium dimer, the singlet-triplet gap of polyacenes and the triplet-quintet gap of \ce{Fe} porphyrin
are taken as showcases in Sec. \ref{sec:pilot_app}, which is followed by concluding remarks in Sec.~\ref{sec:conclusion}.

\section{THEORY}\label{sec:theory}

\subsection{iCI(2)}
The restricted static-dynamic-static (SDS) framework\cite{SDS} for constructing many-electron wave functions
starts with a CAS-based minimal MRCI (denoted as SDSCI) of dimension $3N_P$ for $N_P$ states, viz.
\begin{align}
|\Psi_I\rangle&=\sum_{k=1}^{N_P}|\Psi_k^{(0)}\rangle\bar{C}_{kI}+\sum_{k=1}^{N_P}|\Xi_k^{(1)}\rangle\bar{C}_{(k+N_P)I}
+\sum_{k=1}^{N_P}|\Theta_k^{(2)}\rangle\bar{C}_{(k+2N_P)I},\label{ixc1}\\
|\Psi_k^{(0)}\rangle&=\sum_{|J\nu\rangle \in P}|J\nu\rangle C^{(0)}_{J\nu, k},\\
|\Xi^{(1)}_k\rangle&=Q\frac{1}{E_k^{(0)}-H_0}QH|\Psi^{(0)}_k\rangle=\sum_{|\Phi_{q} \rangle \in Q}|\Phi_{q}\rangle C^{(1)}_{q, k},\label{PTWF1def}\\
|\Theta^{(2)}_k\rangle&=P_s H |\Xi^{(1)}_k\rangle
=\sum_{|J\nu\rangle\in P} |J\nu\rangle C^{(2)}_{J\nu,k}, \label{SecondaryWFdefapp}\\
P&=P_m+P_s,  \quad Q=1-P, \label{Q1def}\\
P_m&=\sum_{k=1}^{N_P} |\Psi_k^{(0)}\rangle\langle\Psi_k^{(0)} |, \quad P_s=P-P_m.\label{Pmdef}
\end{align}
Here, $|\Psi_k^{(0)}\rangle$ is an eigenfunction of the CASCI Hamiltonian $P H P$, while
$|\Xi_k^{(1)}\rangle$ and $|\Theta_k^{(2)}\rangle$
represent the first-order and secondary functions specific to state $k$, respectively.
The expansion coefficients $\bar{\mathbf{C}}$ are determined by the generalized secular equation
\begin{eqnarray}
\bar{\mathbf{H}}\bar{\mathbf{C}}=\bar{\mathbf{S}}\bar{\mathbf{C}}\mathbf{E}, \label{EigenEQ}
\end{eqnarray}
where the reduced Hamiltonian $\bar{\mathbf{H}}$ and metric $\bar{\mathbf{S}}$ have the following structure
\begin{eqnarray}
\bar{\mathbf{H}}&=&\begin{pmatrix}
P_mHP_m& P_mHQ& P_mHP_s\\
QHP_m&QHQ&QHP_s\\
P_sHP_m&P_sHQ&P_sHP_s\end{pmatrix} \label{Hblcok}\\
&=&\begin{pmatrix}E_{k}^{(0)}\delta_{kl} & \langle\Psi_k^{(0)}|H|\Xi_l^{(1)}\rangle &0 \\
 \langle\Xi_l^{(1)}|H|\Psi_k^{(0)}\rangle& \langle\Xi_k^{(1)}|H|\Xi_l^{(1)}\rangle &\langle\Xi_k^{(1)}|H|\Theta_l^{(2)}\rangle\\
 0&\langle\Theta_l^{(2)}|H|\Xi_k^{(1)}\rangle& \langle\Theta_k^{(2)}|H|\Theta_l^{(2)}\rangle\end{pmatrix}, \quad k, l \in[1, N_P],\label{Hmat}\\
\bar{\mathbf{S}}&=&\begin{pmatrix}\delta_{kl} & 0 &0 \\
 0& \langle\Xi_k^{(1)}|\Xi_l^{(1)}\rangle & 0\\
 0&0& \langle\Theta_k^{(2)}|\Theta_l^{(2)}\rangle\end{pmatrix}, \quad k, l \in[1, N_P].
\end{eqnarray}
It follows that the above SDSCI combines the good of perturbation theory [cf. Eq. \eqref{PTWF1def}], intermediate Hamiltonian [cf. Eq. \eqref{SecondaryWFdefapp}]
and configuration interaction [cf. Eq. \eqref{EigenEQ}] in the most economical way:
no matter how many electrons and how many orbitals are to be correlated, only a
$3N_P$-by-$3N_P$ Hamiltonian matrix \eqref{Hmat} is constructed and diagonalized for $N_P$ states.
It has been shown\cite{SDSRev,QUEST4X} that
SDSCI has similar accuracy to ic-MRCISD\cite{XianCI2018} (internally contracted multireference configuration interaction
with singles and doubles), but with a computational cost of just one iteration of the latter.
If the $QHQ$ block of the Hamiltonian matrix \eqref{Hblcok} is replaced with $QH_0Q$, we obtain
SDSPT2\cite{SDS,SDSPT2,SDSPT2s}, a CI-like MRPT2 that treats single and multiple states in the same way.
SDSPT2 is particularly advantageous for situations with multiple nearly degenerate states\cite{SDS,SDSPT2}, for
the existence of secondary functions facilitates relaxations of the reference functions in the presence of dynamic correlation,
as can be seen from the fact that the secondary states $\{|\Theta^{(2)}_k\rangle\}$ arise
from $\{|\Xi_k^{(1)}\rangle\}$ but are projected onto the domain of $\{|\Psi_k^{(0)}\rangle\}$.
On the other hand, if the SDSCI eigennvectors are taken as a new start
and the SDS procedure is repeated until convergence, we obtain iCI\cite{iCI}, which is an exact
solver of FCI (where a $3N_P$-by-$3N_P$ matrix
is constructed and diagonalized in each iteration).
Combined with the selection of configurations, we obtain selected iCI (SiCI).
Here, the following criteria\cite{iCIPT2,iCIPT2New} for the importance measure of individual CSFs are invoked (for a single state for brevity)
\begin{itemize}
	\item [(A)] If $|I\rangle$ is identical with or singly excited from $|J\rangle$, then
	\begin{equation}\label{iCIS}
		\begin{aligned}
			& f\left(|I \mu\rangle,|J\rangle, C_{\min }\right)=\left(\max _\nu\left(\left|H_{\mu \nu}^{I J} C^{(0)}_{J\nu}\right|\right) \geq C_{\min }\right) \quad \text { and } \\
			& \left(\max _\nu\left(\left|\frac{H_{\mu \nu}^{I J} C^{(0)}_{J\nu}}{E^{(0)}-H_{\mu \mu}^{I I}}\right|\right) \geq C_{\min }\right)
		\end{aligned}
	\end{equation}
	\item [(B)] If $|I\rangle$ is doubly excited from $|J\rangle$, then
	\begin{equation}\label{iCID}
		\begin{aligned}
			& \left.f(| I \mu\rangle,|J\rangle, C_{\min }\right)=\left(\max _\nu\left(\left|\tilde{H}^{I J} C^{(0)}_{J\nu}\right|\right) \geq C_{\min }\right) \quad \text { and } \\
			& \quad\left(\max _\nu\left(\left|H_{\mu \nu}^{I J} C^{(0)}_{J\nu}\right|\right) \geq C_{\min }\right) \quad \text { and } \\
			& \quad\left(\max _\nu\left(\left|\frac{H_{\mu \nu}^{I J} C^{(0)}_{J\nu}}{E^{(0)}-H_{\mu \mu}^{I I}}\right|\right) \geq C_{\min }\right)
		\end{aligned}
	\end{equation}
\end{itemize}
Specifically, for case (A), loop over all external CSFs $|I\mu\rangle$ and evaluate $H^{IJ}_{\mu\nu}$ ($=\langle I\mu|H|J\nu\rangle$) for all CSFs $|J\nu\rangle$
belonging to the guess space $P_0[C_{\mathrm{min}}]$
(which provides wave function $|\Psi^{(0)}\rangle = \sum_{|J\nu\rangle \in P_0[C_{\mathrm{min}}]} C^{(0)}_{J\nu} |J\nu\rangle$ with energy $E^{(0)}$).
If $\max_{\nu}(|H^{IJ}_{\mu\nu}C^{(0)}_{J\nu}|)$ is larger than $C_{\mathrm{min}}$ then evaluate $H^{II}_{\mu\mu}$; otherwise discard $|I\mu\rangle$. If $\max_{\nu}(|\frac{H^{IJ}_{\mu\nu}C^{(0)}_{J\nu}}{E^{(0)}-H^{II}_{\mu\mu}}|)$ is larger than $C_{\mathrm{min}}$ then $|I\mu\rangle$ is selected.
As for case (B), only those doubly excited oCFGs $|I\rangle$ with upper bounds $\tilde{H}^{IJ}$\cite{iCIPT2}
larger than $C_{\mathrm{min}}/\max_{\nu} |C_{J\nu}|$ need to be generated
(i.e., those unimportant ones are never touched, as in heat-bath CI\cite{HBCI2016}).
For such $\{|I\rangle\}$, the remaining step is the same as case (A). The so-selected $|I\mu\rangle$ and those in $P_0[C_{\mathrm{min}}]$ form
an expanded space $P_1$, which is diagonalized by iVI\cite{iVI,iVI-TDDFT} to prune away those
CSFs of coefficients smaller in magnitude than $C_{\mathrm{min}}$, thereby reducing $P_1$ to $P[C_{\mathrm{min}}]$.
The latter is taken as the guess $P_0[C_{\mathrm{min}}]$ for the next iteration.
The iterative selection is terminated if the similarity measure
$\frac{\left|P_0\left[C_{\text{min}}\right] \cap P\left[C_{\text{min}}\right]\right|}{\left|P_0\left[C_{\text{min}}\right] \cup P\left[C_{\text{min}}\right]\right|} $
has reached a predefined threshold (e.g., 95\%).

The above SiCI can generate more compact reference spaces than heat-bath CI\cite{HBCI2016} in a more efficient way than CIPSI
(configuration interaction with perturbative selection made iteratively)\cite{CIPSIa}, because both the numerator and denominator
of the ENPT1 amplitude are exploited yet without the expensive summation over reference functions (for more details, see Ref. \citenum{iCIPT2New}). 
The residual dynamic correlation can be estimated with the state-specific ENPT2\cite{Epstein,Nesbet}
\begin{align}
E_{c}^{\mathrm{ENPT2}}[Q,P_m]&=\sum_{|I \mu\rangle \in Q} \frac{\left|\left\langle I \mu|H| \Psi^{(0)}[P_m]\right\rangle\right|^2}
{E^{(0)}-H_{\mu \mu}^{I I}}, \label{eqn:pt2_formula}
\end{align}
where $P_m=P[C_{\mathrm{min}}]$ and $Q=1-P_m$. A pipelined, residue- and constraint-based algorithm has been developed \cite{iCIPT2New,MetaWave}
to evaluate Eq.~\eqref{eqn:pt2_formula} efficiently. The central ingredients of the algorithm include
a memory-efficient algorithm for finding interacting CSF pairs and the tabulated unitary group approach\cite{iCIPT2} for evaluating
their Hamiltonian matrix elements.

The above SiCI+PT2 method is originally termed iCIPT2\cite{iCIPT2}, where the selection is always performed over the full Hilbert space, without
recourse to the concept of active space. When combined with extrapolations, iCIPT2 can provide near-FCI results\cite{benzene}
and is therefore a near-exact method. In the present context, the SiCI+PT2 method is to be dubbed iCI(2), to emphasize
that the selection is to be performed over a large CAS $P$. The $Q$ space in Eq. \eqref{eqn:pt2_formula} then
refers to $P_s=P-P_m$, with $P_m=P[C_{\mathrm{min}}]$ (more precisely, $P_s$ is spanned by single and double excitations from $P_m$
and is thus generally smaller than $P-P_m$. However, we ignore this subtlety, for it does not affect the result). 
Accordingly, the energy
$E^{\mathrm{iCI(2)}}[P_s, P_m]=E^{(0)}[P_m]+E^{\mathrm{ENPT2}}_c[P_s, P_m]$ by iCI(2) is an approximation to the corresponding CASCI energy.
We also have iCISCF(2)\cite{iCISCF} as an approximation to CASSCF, where both the CI coefficients and orbitals are optimized iteratively.

\subsection{iCI-MRPT2}
Having determined an adaptive PAS $P_m$, which provides the zeroth-order wave function
\begin{align}
P_mHP_m |\Psi^{(0)}[P_m]\rangle&=|\Psi^{(0)}[P_m]\rangle E^{(0)}[P_m],\\
|\Psi^{(0)}[P_m]\rangle&=\sum_{|J\nu\rangle \in P_m}|J\nu\rangle C^{(0)}_{J\nu},
\end{align}
as well as the (inner-space) second-order energy $E^{\mathrm{ENPT2}}_c[P_s, P_m]$ in Eq. \eqref{eqn:pt2_formula},
the remaining task is to account for the dynamic correlation correction of the external space $Q$ ($=1-P=1-P_m-P_s$) to the variational space $P_m$
[NB: the $P_m$ here is spanned by individually selected CSFs and is hence different from that in Eq. \eqref{Pmdef}]. This requires specification of the zeroth-order Hamiltonian
and XFs. For the latter, a partial contraction scheme is adopted here to avoid higher than 2-RDMs. Specifically,
the XFs belonging to $Q_c= V_E\oplus V_G\oplus V_H$ (see Table \ref{table:FOIS_MRPT2}) are fully internally contracted,
whereas those belonging to $Q_u=Q-Q_c= V_A+V_B+V_C+V_D+V_F$ are kept uncontracted.
Since $Q_c$ lies energetically well above $P_m$, the contraction should
lead to very small errors but eliminate the high scaling with the number of inactive orbitals.
In accordance with this design, we consider the following zeroth-order Hamiltonian
\begin{equation}
	H_0=P_mHP_m + \sum_{|I\mu\rangle \in P_s} |I\mu\rangle H^{II}_{\mu\mu} \langle I\mu | + Q_c H^D  Q_c
	+ Q_u H^U_0 Q_u,\label{H0def}
\end{equation}
which gives rise to the following energy up to second order
\begin{align}
	E_{\text{tot}}[Q_t, P_m]&= E^{\mathrm{iCI(2)}}[P_s, P_m] + E^{\mathrm{MRPT2}}_c[Q_c, P_m] + E^{\mathrm{MRPT2}}_c[Q_u, P_m],\label{iCIMRPT2Ene}\\
Q_t&=P_s+Q_c+Q_u,\nonumber\\
E^{\mathrm{iCI(2)}}[P_s, P_m]&=E^{(0)}[P_m]+E^{\mathrm{ENPT2}}_c[P_s, P_m].
\end{align}
Specifically, ENPT2 is employed for the $P_s$ space for an easy extrapolation to zero $C_{\mathrm{min}}$,
so as to make the iCI(2)/iCISCF(2) energy close to the CASCI/CASSCF value. To ensure
orbital invariance within the contracted space $Q_c$, the Dyall Hamiltonian\cite{DyallHmat} is employed
\begin{align}
H^D             &= H^D_{\mathrm{I}}+H^D_{\mathrm{A}},\label{H0D}\\
H^D_{\mathrm{I}}&= f_{ii} E_{ii}+ f_{aa} E_{aa}+C_D,\\
H^D_{\mathrm{A}}&=f^c_{tu}E_{tu} +\frac{1}{2} (tu|vw)e_{tu,vw},\\
f_{pq}          &=f^c_{pq}+[(pq|tu)-\frac{1}{2}(pu|tq)]D_{tu},\label{fmat}\\
f^c_{pq}        &=h_{pq}+ 2(pq|ii)-(pi|iq),\\
D_{rs}          &=\langle \Psi^{(0)}[P_m]|E_{rs}|\Psi^{(0)}[P_m]\rangle,\\
C_D             &=E^c-2f_{ii},\quad E^c= h_{ii} + f^c_{ii},\label{DyallC}
\end{align}
The particular choice of the constant $C_D$ \eqref{DyallC} is to ensure that $PH^DP$ has the same spectrum as $PHP$.
As for $H_0^U$ in Eq. \eqref{H0def}, it can be $H^D$ (denoted as iCI-NEVPT2)
or the generalized Fock operator
\begin{align}
H_0^U&=F+ C_F, \label{H0UF}\\
  F&=f_{pq}E_{pq},\label{eqn:gfock}\\
  C_F&=E^{(0)}[P_m]-\langle \Psi_0[P_m]|F|\Psi_0[P_m]\rangle.
\end{align}
Here, the constant $C_F$ is to align the zeroth-order energies of individual CSFs in $Q_u$ to $E^{(0)}[P_m]$.
This variant is to be denoted as iCI-MRMP2, although the term MRMP2 has been used\cite{HiraoMRMP2} in a different context
(i.e., the present, last term in Eq. \eqref{H0def},
$|I\mu\rangle\langle I\mu|F|J\nu\rangle\langle J\nu|$,
is different from the diagonal form $|I\mu\rangle\langle I\mu|F|I\mu\rangle\langle I\mu|$ in Ref. \citenum{HiraoMRMP2}).
Additional variants of $H_0^U$ can be introduced. For instance, only the class-diagonal part $F_{CD}$ of $F$ \eqref{eqn:gfock}
may be considered
\begin{align}
H_0^U&= F_{CD} + C_{CD},\label{H0UFD}\\
F_{CD}&=f_{ij}E_{ij} + f_{uv}E_{uv}+f_{ab}E_{ab},\label{FcdOper}\\
C_{CD}&=  E^{(0)}[P_m]-\langle \Psi^{(0)}[P_m]|F_{CD}|\Psi^{(0)}[P_m]\rangle.
\end{align}
This variant is to be dubbed iCI-MRMP2D. Likewise, adding the class-off-diagonal part $F_{CO}=F-F_{CD}$ of $F$
\begin{align}
F_{CO}&=\{f_{ui}E_{ui} + f_{ai}E_{ai}+f_{av}E_{av}\}+\mathrm{h.c.} \label{FcoOper}
\end{align}
to the Dyall Hamiltonian gives rise to an extended Dyall Hamiltonian for $H_0^U$
\begin{equation}
	H_0^U = H^D + F_{CO}.\label{H0UDE}
\end{equation}
This variant can obviously be denoted as iCI-NEVPT2E. Finally,
the following diagonal forms for $H_0^U$ can be introduced just for comparison,
\begin{align}
H_0^U &= \sum_{|I\mu\rangle \in Q_u} |I\mu\rangle H^{II}_{\mu\mu} \langle I\mu |,\label{H0UENPT}\\
H_0^U &= \sum_{|I\mu\rangle \in Q_u} |I\mu\rangle (H^{\text{D}}_0)^{II}_{\mu\mu} \langle I\mu |,\label{H0UdiaD}
\end{align}
which are denoted as iCI-MRENPT2 and iCI-NEVPT2D, respectively.

The first-order amplitudes $\mathbf{t}$ are determined by the following generic linear equations
\begin{align}
\sum_{|\chi_l\rangle} \langle\chi_k|Q_x(H_0-E^{(0)})Q_x|\chi_l\rangle t_l=-\langle \chi_k|Q_xV|\Psi^{(0)}[P_m]\rangle,
\quad Q_x= P_s, Q_c, Q_u^c,\label{LinearEQ0}
\end{align}
which can be written in matrix form
\begin{align}
(\mathbf{H}_0-E^{(0)}\mathbf{M})\mathbf{t}=-\mathbf{V}.\label{LinearEQ1}
\end{align}
Here, $\{|\chi_k\rangle\}\in Q_x$ denote the XFs and $M_{kl}=\langle\chi_k|\chi_l\rangle$. Since all reference functions $\{|\Phi_\mu\rangle\}$
share the same occupation pattern for the inactive (core or virtual) orbitals,
the matrix elements $\langle \Phi_\mu|\hat{E}_M^\dag H_0\hat{E}_N|\Phi_\nu\rangle$ are nonvanishing only when $\hat{E}_M^\dag H_0\hat{E}_N$
have the same number of core/virtual creation and annihilation operators, so that the second-quantized operators after the Wick contractions
involve only active orbitals (NB: $\hat{E}_N$ refers to one of the eight classes of excitation operators in Table \ref{table:FOIS_MRPT2}). A direct deduction is that the particle/hole-number-conserving operators $H^D$ and $F_{CD}$ can only couple XFs belonging to the same $Q_l^{(k)}$ subspace, whereas the particle/hole-number-non-conserving operator $F_{CO}$ can only couple XFs belonging to $Q_l^{(k)}$ and those belonging to $Q_m^{(k)}$ with $m\ne l$ (for more details, see Table \ref{table:coupling_PQ}).
Therefore, iCI-MRMP2/NEVPT2E can be viewed as the coupled version of iCI-MRMP2D/NEVPT2, just like that CASPT2 is the coupled version of CASPT2D\cite{CASPT2Rev}.
With the first-order amplitudes $\mathbf{t}$, the second-order correlation energy reads
\begin{align}
E^{\mathrm{MRPT2}}_c[Q_x,P_m]=\mathbf{V}^\dag \mathbf{t}, \quad Q_x= P_s, Q_c, Q_u^c.\label{MRMPT2Ec}
\end{align}

For clarity, the various variants of iCI-MRPT2 defined by different $H_0^U$ are documented in Table \ref{iCI-MRPT2}.


\begin{table}[htbp]
	\centering
	\caption{Terms$^*$ of $H_0^U$ that couple the $P$- and $Q_u$-space functions. }
	\label{table:coupling_PQ}
	\begin{threeparttable}
		\begin{tabular}{c|cccccc} \toprule
			               & $P$ & $Q_{,i}^{(1)}$ & $Q_{,ij}^{(2)}$ & $Q_{a}^{(-1)}$ & $Q_{a,i}^{(0)}$  & $Q_{ab}^{(-2)}$ \\  \toprule
			 $P$           & $F_{CD}/H^D$ & $F_{\mathrm{co}}$ &   & $F_{\mathrm{ov}}$ & $F_{\mathrm{cv}}$ &   \\
			$Q_{,i}^{(1)}$ & $F_{\mathrm{oc}}$ & $F_{CD}/H^D$ & $F_{\mathrm{co}}$ &   & $F_{\mathrm{ov}}$ &   \\
			$Q_{,ij}^{(2)}$&& $F_{\mathrm{oc}}$ & $F_{CD}/H^D$ &   &   &   \\
			$Q_{a}^{(-1)}$ & $F_{\mathrm{vo}}$ &   &   & $F_{CD}/H^D$ & $F_{\mathrm{co}}$ & $F_{\mathrm{ov}}$ \\
			$Q_{a,i}^{(0)}$& $F_{\mathrm{vc}}$ & $F_{\mathrm{vo}}$ &   & $F_{\mathrm{oc}}$ & $F_{CD}/H^D$ &   \\
			$Q_{ab}^{(-2)}$&&   &   & $F_{\mathrm{vo}}$ &   & $F_{CD}/H^D$ \\ \bottomrule
		\end{tabular}
		\begin{tablenotes}
			\item[*] $F_{\mathrm{co}}$, $F_{\mathrm{ov}}$ and $F_{\mathrm{cv}}$ are the core-active, active-virtual 
and core-virtual parts of $F_{CO}$, respectively. 
		\end{tablenotes}
	\end{threeparttable} 
\end{table}

\begin{table}[htbp]
	\centering
\small
	\caption{Variants of iCI-MRPT2 defined by different $H_0^U$ in Eq. \eqref{H0def}.
	}
	\begin{threeparttable}
		\begin{tabular}{lcccccccccccccccc} \toprule
		\multicolumn{1}{l}{iCI-MRPT2}
		& \multicolumn{1}{c}{$H_0^U$} \\ \toprule
        iCI-MRMP2D   & Eq. \eqref{H0UFD} & \\
        iCI-MRMP2    & Eq. \eqref{H0UF}  & \\
        iCI-MRMP2*\tnote{a}    & Eq. \eqref{H0UF}  & \\
        iCI-NEVPT2D  & Eq. \eqref{H0UdiaD} &\\
        iCI-NEVPT2   & Eq. \eqref{H0D}   & \\
        iCI-NEVPT2E  & Eq. \eqref{H0UDE} & \\
        iCI-NEVPT2E*\tnote{a}  & Eq. \eqref{H0UDE} & \\
        iCI-MRENPT2  & Eq. \eqref{H0UENPT} &\\
        \bottomrule
		\end{tabular}
\begin{tablenotes}
\item[a] With LCUT (see Sec. \ref{LCUT}).
\end{tablenotes}
	\end{threeparttable}\label{iCI-MRPT2}
\end{table}

\subsection{Size-consistency of iCI-MRPT2}\label{SecSizeCon}

As shown recently, stringent conditions have to be invoked for a PAS-based MRPT2 to be size-extensive\cite{PASPT2,PASPT2New}.
Therefore, the size-extensivity of iCI-MRPT2 is out of the question due to lack of such conditions. Nevertheless, some variants of iCI-MRPT2 can still be
size-consistent, meaning\cite{Size-consistency} that the calculated energy for a compound system AB composed of
two physically separated, non-interacting subsystems A and B is precisely equal to the sum of the energies
of A and B calculated separately. To see this,  we start with a generic zeroth-order problem
\begin{align}
H^{(0)}&=PHP+QW^\prime Q,\quad Q=1-P \nonumber\\
&=\sum_{|p\rangle,|p^\prime\rangle \in P}|p\rangle\langle p|H|p^\prime\rangle\langle p^\prime|+QW^\prime Q\nonumber\\
&=\sum_{|K\rangle\in P}|K\rangle E_K^{(0)}\langle K| +QW^\prime Q, \label{H0definition}\\
W^\prime&=W+C_W, \quad C_W=E_0^{(0)}-\langle 0|W|0\rangle
\end{align}
where the constant $C_W$ is to align the zeroth-order energies of the XFs to $E_0^{(0)}$. Both the Hamiltonian $H$ and the operator $W$ are additively separable
\begin{align}
H_{AB}=H_A+H_B, \quad W_{AB}=W_A+W_B,
\end{align}
whereas the zeroth-order wave functions $\{|K\rangle=|p\rangle C_{pK}^{(0)}\}$ (eigenfunctions of $PHP$) are multiplicatively separable
\begin{align}
|K_{AB}\rangle=|K_A\rangle \wedge |K_B\rangle=|K_A K_B\rangle,
\end{align}
so that
\begin{align}
E_{K_{AB}}^{(0)}&=\langle K_{AB}|H_{AB}|K_{AB}\rangle \nonumber\\
&=\langle K_{A}|H_{A}|K_{A}\rangle+\langle K_{B}|H_{B}|K_{B}\rangle\nonumber\\
&=E_{K_A}^{(0)}+E_{K_B}^{(0)}.\label{E0AB}
\end{align}
By virtue of the relations
\begin{align}
P_{AB}&=P_A\otimes P_B,\\
Q_{AB}&=1_A\otimes 1_B-P_{AB}\nonumber\\
&=Q_A\otimes P_B+P_A\otimes Q_B+Q_A\otimes Q_B\nonumber\\
&=Q_A\otimes 1_B+P_A\otimes Q_B\nonumber\\
&=1_A\otimes Q_B+Q_A\otimes P_B,\label{QABDeompos}
\end{align}
the zeroth-order Hamiltonian $H^{(0)}_{AB}$ for the compound system AB can be decomposed as
\begin{align}
H^{(0)}_{AB}&=\tilde{H}^{(0)}_{AB}+\bar{H}^{(0)}_{AB},\label{H0decomposition}\\
\tilde{H}^{(0)}_{AB}&=H^{(0)}_A\otimes 1_B+ 1_A\otimes H^{(0)}_B, \label{SepH0}\\
H^{(0)}_Z&=\sum_{|K_Z\rangle\in P_Z} |K_Z\rangle E_{K_Z}^{(0)}\langle K_Z|+Q_ZW_Z^\prime Q_Z,\quad Z = A, B,\\
W_Z^\prime&=W_Z+C_Z,\quad C_Z=E_{0_Z}^{(0)}-\langle 0_Z|W_Z|0_Z\rangle,\\
\bar{H}^{(0)}_{AB}&=\bar{H}^{(0)o}_{AB}+\bar{H}^{(0)d}_{AB},\label{NonsepH0}\\
\bar{H}^{(0)o}_{AB}&=(P_A W_A Q_A+\mathrm{h.c.})\otimes Q_B+Q_A\otimes (P_B W_B Q_B+\mathrm{h.c.}),\\
\bar{H}^{(0)d}_{AB}&=P_A(W_A^\prime-H_A)P_A\otimes Q_B+Q_A\otimes P_B(W_B^\prime-H_B)P_B,
\end{align}
where $\tilde{H}^{(0)}_{AB}$ and $\bar{H}^{(0)}_{AB}$ are the additively separable and non-separable parts of $H^{(0)}_{AB}$, respectively.
Recognizing that both $\tilde{H}^{(0)}_{AB}$ and $\bar{H}^{(0)d}_{AB}$ are even (diagonal), whereas
$\bar{H}^{(0)o}_{AB}$ is odd (off-diagonal), the matrix elements of $H^{(0)}_{AB}-E^{(0)}_{0_{AB}}$
over the  XFs $\{|q_AK_B\rangle\}$, $\{|K_Aq_B\rangle\}$ and $\{|q_Aq_B\rangle\}$ (NB: $\{q_Z\}$ represent the (internally contracted or uncontracted)
XFs on subsystem Z) can readily be calculated as
\begin{align}
(A_{11})_{q_AK_B, q_A^\prime K_B^\prime}&=\langle q_AK_B|H^{(0)}_{AB}-E^{(0)}_{0_{AB}}|q_A^\prime K_B^\prime\rangle\nonumber\\
&=\delta_{K_BK_B^\prime}\langle q_A|W_A^\prime-E_{0_A}^{(0)}+E_{K_B}^{(0)}-E_{0_B}^{(0)}|q_A^\prime\rangle\nonumber\\
&+\delta_{q_Aq_A^\prime}\langle K_B|W_B^\prime-E_{K_B}^{(0)}|K_B^\prime\rangle,\label{A11}\\
(A_{12})_{q_AK_B, K_A^\prime q_B^\prime}&=\langle q_AK_B|H^{(0)}_{AB}-E^{(0)}_{0_{AB}}|K_A^\prime q_B^\prime\rangle=0,\label{A12}\\
(A_{13})_{q_AK_B,q_A^\prime q_B^\prime}&=\langle q_AK_B|H^{(0)}_{AB}-E^{(0)}_{0_{AB}}|q_A^\prime q_B^\prime\rangle\nonumber\\
&=\delta_{q_Aq_A^\prime}\langle K_B|W_B^\prime|q_B^\prime\rangle,\label{A13}\\
(A_{21})_{K_Aq_B, q_A^\prime K_B^\prime}&=\langle K_Aq_B|H^{(0)}_{AB}-E^{(0)}_{0_{AB}}|q_A^\prime K_B^\prime\rangle=0,\label{A21}\\
(A_{22})_{K_Aq_B, K_A^\prime q_B^\prime}&=\langle K_Aq_B|H^{(0)}_{AB}-E^{(0)}_{0_{AB}}|K_A^\prime q_B^\prime\rangle\nonumber\\
&=\delta_{K_AK_A^\prime}\langle q_B|W_B^\prime-E_{0_B}^{(0)}+E_{K_A}^{(0)}-E_{0_A}^{(0)}|q_B^\prime\rangle\nonumber\\
&+\delta_{q_Bq_B^\prime}\langle K_A|W_A^\prime-E_{K_A}^{(0)}|K_A^\prime\rangle,\label{A22}\\
(A_{23})_{K_Aq_B, q_A^\prime q_B^\prime}&=\langle K_Aq_B|H^{(0)}_{AB}-E^{(0)}_{0_{AB}}|q_A^\prime q_B^\prime\rangle\nonumber\\
&=\delta_{q_Bq_B^\prime}\langle K_A|W_A^\prime|q_A^\prime\rangle,\label{A23}\\
(A_{31})_{q_Aq_B, q_A^\prime K_B^\prime}&=\langle q_Aq_B|H^{(0)}_{AB}-E^{(0)}_{0_{AB}}|q_A^\prime K_B^\prime\rangle\nonumber\\
&=\delta_{q_Aq_A^\prime}\langle q_B|W_B^\prime|K_B^\prime\rangle,\label{A31}\\
(A_{32})_{q_Aq_B, K_A^\prime q_B^\prime}&=\langle q_Aq_B|H^{(0)}_{AB}-E^{(0)}_{0_{AB}}|K_A^\prime q_B^\prime\rangle\nonumber\\
&=\delta_{q_Bq_B^\prime}\langle q_A|W_A^\prime|K_A^\prime\rangle,\label{A32}\\
(A_{33})_{q_Aq_B, q_A^\prime q_B^\prime}&=\langle q_Aq_B|H^{(0)}_{AB}-E^{(0)}_{0_{AB}}|q_A^\prime q_B^\prime\rangle\nonumber\\
&=\delta_{q_Aq_A^\prime}\langle q_B|W_B^\prime|q_B^\prime\rangle+\delta_{q_Bq_B^\prime}\langle q_A|W_A^\prime|q_A^\prime\rangle.\label{A33}
\end{align}
Moreover, we have the following expressions for the perturbation operator $V_{AB}$ ($=H_{AB}-H^0_{AB}$):
\begin{align}
(V_{10})_{q_AK_B, 0_A 0_B}&=\langle q_AK_B|H_{AB}|0_A 0_B\rangle\nonumber\\
&=\delta_{K_B,0_B}\langle q_A|H_A|0_A\rangle,\label{V10}\\
(V_{20})_{K_Aq_B, 0_A 0_B}&=\langle K_Aq_B|H_{AB}|0_A 0_B\rangle\nonumber\\
&=\delta_{K_A,0_A}\langle q_B|H_B|0_B\rangle,\label{V20}\\
(V_{30})_{q_Aq_B, 0_A 0_B}&=\langle q_Aq_B|H_{AB}|0_A 0_B\rangle=0. \label{V30}
\end{align}
At this stage, it is important to point out that the above expressions for $A_{ij}$  and $V_{i0}$ can
also be obtained by using the XFs $\{|q_A p_B\rangle\}$ and $\{|p_A q_B\rangle\}$ for the $Q_A\otimes P_B$ and $P_A\otimes Q_B$
spaces, respectively, followed by transforming $\{|p_A\rangle\}$ to $\{|K_A\rangle\}$ with $\{C_{p_AK_A}^{(0)}\}$ and
$\{|p_B\rangle\}$ to $\{|K_B\rangle\}$ with $\{C_{p_BK_B}^{(0)}\}$. It is of interest to see that Eq. \eqref{V10}/\eqref{V20}
dictates that $|K_B\rangle$ ($|K_A\rangle$) in the elements of $A_{11}$ and $A_{13}$ ($A_{22}$ and $A_{23}$) can only be $|0_B\rangle$ ($|0_A\rangle$), so that
\begin{align}
(A_{11})_{q_A0_B, q_A^\prime K_B^\prime}
&=\delta_{0_BK_B^\prime}\langle q_A|W_A^\prime-E_{0_A}^{(0)}|q_A^\prime\rangle\nonumber\\
&+\delta_{q_Aq_A^\prime}\langle 0_B|W_B^\prime-E_{0_B}^{(0)}|K_B^\prime\rangle,\label{A110}\\
(A_{13})_{q_A0_B,q_A^\prime q_B^\prime}
&=\delta_{q_Aq_A^\prime}\langle 0_B|W_B^\prime|q_B^\prime\rangle,\label{A130}\\
(A_{22})_{0_Aq_B, K_A^\prime q_B^\prime}
&=\delta_{0_AK_A^\prime}\langle q_B|W_B^\prime-E_{0_B}^{(0)}|q_B^\prime\rangle\nonumber\\
&+\delta_{q_Bq_B^\prime}\langle 0_A|W_A^\prime-E_{0_A}^{(0)}|K_A^\prime\rangle,\label{A220}\\
(A_{23})_{0_Aq_B, q_A^\prime q_B^\prime}
&=\delta_{q_Bq_B^\prime}\langle 0_A|W_A^\prime|q_A^\prime\rangle.\label{A230}
\end{align}
The linear equations for the first-order amplitudes $\mathbf{t}$ then read
\begin{align}
\begin{pmatrix}
(A_{11})_{q_A0_B, q_A^\prime K_B^\prime}&0                                       &(A_{13})_{q_A0_B,q_A^\prime q_B^\prime}\\
0                                       &(A_{22})_{0_Aq_B, K_A^\prime q_B^\prime}&(A_{23})_{0_Aq_B, q_A^\prime q_B^\prime}\\
(A_{31})_{q_Aq_B, q_A^\prime K_B^\prime}&(A_{32})_{q_Aq_B, K_A^\prime q_B^\prime}&(A_{33})_{q_Aq_B, q_A^\prime q_B^\prime}
\end{pmatrix}
\begin{pmatrix}
t_{q_A^\prime K_B^\prime}\\
t_{K_A^\prime q_B^\prime}\\
t_{q_A^\prime q_B^\prime}\\
\end{pmatrix}
=-
\begin{pmatrix}
(V_{10})_{q_A0_B, 0_A 0_B}\\
(V_{20})_{0_Aq_B, 0_A 0_B}\\
0\\
\end{pmatrix}.\label{ABteq}
\end{align}

After the above general formulation, specific applications can now be made. Firstly,
if the $W$ operator in $H^0$ \eqref{H0definition} is chosen to be the Dyall Hamiltonian $H^D$ (which has the same $P$-space states $\{|K\rangle\}$ as H),
then $C_W$ and $\bar{H}^{(0)}_{AB}$ (and hence $A_{i3}$ and $A_{3i}$) vanish. Moreover, $A_{11}$ \eqref{A110} and $A_{22}$ \eqref{A220}
reduce to
\begin{align}
(A_{11})_{q_A0_B, q_A^\prime 0_B}&=\langle q_A|H^D_A-E_{0_A}^{(0)}|q_A^\prime\rangle,\label{A110HD}\\
(A_{22})_{0_Aq_B, 0_A q_B^\prime}&=\langle q_B|H^D_B-E_{0_B}^{(0)}|q_B^\prime\rangle,\label{A220HD}
\end{align}
so that Eq. \eqref{ABteq} are decoupled precisely to those for the separated A and B. Consequently, iCI-NEVPT2 is size-consistent,
as can be expected. Secondly, if $W=H^D+F_{CO}$, then $A_{11}$ \eqref{A110} and $A_{22}$  \eqref{A220} also become the same as
those for the separated A and B, respectively, viz.
\begin{align}
(A_{11})_{q_A0_B, q_A^\prime 0_B}&=\langle q_A|H^D_A+F_{CO}-E_{0_A}^{(0)}|q_A^\prime\rangle,\label{A110HDFCO}\\
(A_{22})_{0_Aq_B, 0_A q_B^\prime}&=\langle q_B|H^D_B+F_{CO}-E_{0_B}^{(0)}|q_B^\prime\rangle.\label{A220HDFCO}
\end{align}
This is because the class-off-diagonal operator $F_{CO}$ introduces changes
in the number of active electrons, so that $\langle K|F_{CO}|K^\prime\rangle\equiv 0$.
However, the $A_{i3}$ and $A_{3i}$ terms no longer vanish in this case. As a result, iCI-NEVPT2E is
not size-consistent. Nevertheless, if the source-free collective double excitations (CDE) $\{|q_Aq_B\rangle\}$ (singles on A \emph{and} singles on B)
are screened out, so as to make $A_{i3}$ and $A_{3i}$ disappear, 
iCI-NEVPT2E will become size-consistent, as confirmed by the results shown in Table \ref{table:size_consistency_be4}.
Thirdly, if $W=F$ or $F_{CD}$, the second terms in
$A_{11}$ \eqref{A110} and $A_{22}$ \eqref{A220} do not vanish in general, so that
$A_{11}$ and $A_{22}$ are different from those for the separated A and B, respectively. Consequently,
both iCI-MRMP2 and iCI-MRMP2D are not size-consistent, even if the CDEs are screened out (see Table \ref{table:size_consistency_be4}).
The same applies also to CASPT2\cite{CASPT2Rev}, thereby correcting our previous opposite statement on this point\cite{PASPT2New}. 
For comparison, the situation is different for PASPT2\cite{PASPT2,PASPT2New}, where the second terms in $A_{11}$ \eqref{A110} and $A_{22}$ \eqref{A220} do vanish (because
$|0_A\rangle$ and $|0_B\rangle$ therein are single determinants), so that PASPT2 can be made numerically size-consistent 
(by screening out the CDEs) even when the non-separable projected Fock operator is used as the zeroth-order Hamiltonian (for more details, see Ref. \citenum{PASPT2arXiv}).  

The above analysis does not apply to iCI-MRENPT2 and iCI-NEVPT2D for two reasons: (1) the only proper XFs are
$\{|q_A p_B\rangle\}$, $\{|p_A q_B\rangle\}$ and $\{|q_A q_B\rangle\}$ for the $Q_A\otimes P_B$, $P_A\otimes Q_B$
and $Q_A\otimes Q_B$ spaces, respectively; (2) the $H^{(0)}_{AB}$ operator cannot be decomposed as in Eq. \eqref{H0decomposition}.
Nevertheless, it can readily be checked that neither the numerators nor the denominators in the amplitudes of iCI-MRENPT2/NEVPT2D
are additively separable. For instance,
\begin{align}
\langle q_Ap_B|H_{AB}|0_A0_B\rangle&=\langle q_A|H_{A}|0_A\rangle\langle p_B|0_B\rangle\nonumber\\
&\ne\langle q_A|H_{A}|0_A\rangle,\\
E^{(0)}_{0_{AB}} - \langle q_Ap_B|H_{AB}|q_Ap_B\rangle&=E^{(0)}_{0_{A}}-\langle q_A|H_A|q_A\rangle\nonumber\\
&+E^{(0)}_{0_{B}}-\langle p_B|H_B|p_B\rangle\nonumber\\
&\ne E^{(0)}_{0_{A}}-\langle q_A|H_A|q_A\rangle.
\end{align}
Therefore, both iCI-MRENPT2 and iCI-NEVPT2D are not size-consistent, see again Table \ref{table:size_consistency_be4}.
Note in passing that, had the XFs  $\{|q_A K_B\rangle\}$, $\{|K_A q_B\rangle\}$ and $\{|q_A q_B\rangle\}$ been used,
both iCI-MRENPT2 and iCI-NEVPT2D would be size-consistent. However, this is not a practical choice
for multireference calculations across molecular geometries. Nevertheless, it can be deduced that both
SR-ENPT2 is size-consistent for closed-shell fragments (see Table \ref{table:size_consistency_be4}), whereas unrestricted Hartree-Fock (UHF)-based
SR-ENPT2 is size-consistent for open-shell fragments, just like MP2 and UHF-MP2, respectively.
SR-NEVPT2 is actually identical to MP2.

\begin{table}[htbp]
	\small
	\centering
	\caption{Size-consistency errors (in m$E_h$) of iCI-MRPT2s for the ground state energy of two non-interacting \ce{Be2} molecules with the cc-pVDZ basis set.
		The error is defined as $E(\ce{Be4}) - E_1(\ce{Be2})-E_2(\ce{Be2})$, where the two \ce{Be2} fragments have bond lengths of 2.4 and 2.6\,\AA, respectively.
		Canonical CASSCF(8,8) orbitals are employed.
	}
	\begin{threeparttable}
		\begin{tabular}{lSS} \toprule
		\multicolumn{1}{c}{Method}
		& \multicolumn{1}{c}{with CDEs\tnote{a}}
		& \multicolumn{1}{c}{without CDEs\tnote{a}} \\\toprule
		iCI-MRMP2D          & -1.136              & -1.136 \\
		iCI-MRMP2*\tnote{b} & -1.107              & -1.107 \\
		iCI-MRMP2           & -1.171              & -1.146 \\		
        iCI-NEVPT2          & \multicolumn{1}{c}{$<10^{-7}$\tnote{d}} & \multicolumn{1}{c}{$<10^{-7}$\tnote{d}} \\
        iCI-NEVPT2E*\tnote{c}&\multicolumn{1}{c}{$<10^{-7}$\tnote{d}} & \multicolumn{1}{c}{$<10^{-7}$\tnote{d}} \\	
		iCI-NEVPT2E         & -0.039              & \multicolumn{1}{c}{$<10^{-7}$\tnote{d}} \\
        iCI-NEVPT2D         &  1.899              &  1.899 \\
        iCI-MRENPT2         &  3.000              &  3.000 \\
        SR-ENPT2\tnote{e}   & \multicolumn{1}{c}{$<10^{-7}$\tnote{d}} &  \multicolumn{1}{c}{$<10^{-7}$\tnote{d}}    \\
SR-NEVPT2/MP2\tnote{e}      & \multicolumn{1}{c}{$<10^{-7}$\tnote{d}} & \multicolumn{1}{c}{$<10^{-7}$\tnote{d}}   \\
        \bottomrule
		\end{tabular}
		\begin{tablenotes}
            \item[a] Collective double excitations (singles on A \emph{and} singles on B).
			\item[b] iCI-MRPT2 with LCUT.
			\item[c] iCI-NEVPT2 with LCUT.
			\item[d] Zero within numerical accuracy.
            \item[e] With canonical Hartree-Fock orbitals.
		\end{tablenotes}
	\end{threeparttable}\label{table:size_consistency_be4}
\end{table}

\section{Simplifications of iCI-MRPT2}
\subsection{Selection of $Q_u$ and extrapolation}\label{SecSelection}
The uncontracted XFs (i.e., all CSFs of the singly and doubly excited oCFGs) scale as high as $|P_m| N_a^3 N_c$,
which becomes prohibitively large for realistic systems. However, only
a relatively small fraction of such XFs has discernible contributions
to the second-order correlation energy. This suggests that a selection can be performed to identify and retain only the most important XFs,
as done by Grimme and Waletzke\cite{MRMP2s}. In the present context, this
can in principle proceed in the same way as SiCI.
It is just that $H_0^U$ should be used in place of $H$ in Eqs. \eqref{iCIS} and \eqref{iCID},
and meanwhile $C_{\mathrm{min}}$ is replaced with another threshold, $Q_{\mathrm{min}}$.
However, since the contributions of those unimportant CSFs (which span the $Q_u^u$ portion of $Q_{u}$) to the correlation energy (denoted as
$E^{\mathrm{ENPT2}}_c[Q_u^u,P_m]$) are to be retained (instead of discarded),
we decide to take the full ENPT1 coefficients (which are more expensive than the iCI criteria
in Eqs. \eqref{iCIS} and \eqref{iCID}, but allows an easy accumulation of the contributions to $E^{\mathrm{ENPT2}}_c[Q_u^u,P_m]$)
as an importance measure, just like CIPSI\cite{CIPSIa}. More details on the selection are postponed to Sec. \ref{SecQuSelect}. Those
selected CSFs (which span the $Q_u^c$ portion of $Q_{u}$)
will enter the linear equations \eqref{LinearEQ1} and give the correlation energy $E^{\mathrm{MRPT2}}_c[Q_u^c,P_m]$.
For a given $C_{\mathrm{min}}$, the sum of $E^{\mathrm{MRPT2}}_c[Q_u^c,P_m]$ and $E^{\mathrm{ENPT2}}_c[Q_u^u,P_m]$
will approach $E^{\mathrm{MRPT2}}_c[Q_u,P_m]$ in Eq. \eqref{iCIMRPT2Ene} in the limit of $Q_{\mathrm{min}}\rightarrow 0$,
provided that such extrapolation is performed separately for each subspace of $Q_u$.
However, there exists a caveat: when the system is composed of several non-interacting fragments,
the $Q_u^c$ space in the limit $Q_{\text{min}} = 0$ does not necessarily coincide with the full $Q_u$ space,
so that the sum of $E^{\mathrm{MRPT2}}_c[Q_u^c,P_m]$ and $E^{\mathrm{ENPT2}}_c[Q_u^u,P_m]$ would not be equal to $E^{\mathrm{MRPT2}}_c[Q_u,P_m]$.
This has to do with the size-consistency issue of iCI-MRPT2 discussed in Sec. \ref{SecSizeCon}.
The complete iCI‑MRPT2 algorithm is summarized in Algorithm~\ref{algo:iCI_MRPT2}.

\begin{algorithm}[H]
	\caption{Algorithm of iCI-MRPT2 with $C_{\text{min}}$ and $Q_{\text{min}}$ extrapolations}\label{alg:mrpt2}
	\begin{algorithmic}[1]
		\Require Arrays Cmin\_array, Qmin\_array; variational space $P_m$, wave function $|\Psi^{(0)}[P_m]\rangle$
                 and energy $E^{(0)}[P_m]$ for each $C_{\text{min}}$; zeroth‑order Hamiltonian $H_0$
		\Ensure Extrapolated iCI-MRPT2 total energy $E_{\text{tot}}$
		\For    {$C_{\text{min}} \in \text{Cmin\_array}$}
		\State  $E_c^{\mathrm{ENPT2}}[P_s, P_m] \gets$ ENPT2 correction from $P_s=P-P_m$
		\State  $E_c^{\mathrm{MRPT2}}[Q_c, P_m] \gets$ NEVPT2 correction from $Q_c$ ($=V_E\cup V_G\cup V_H$)
		\For    {$Q_{\text{min}} \in \text{Qmin\_array}$}
		\State  Split $Q_u$ into $Q_u^c$ and $Q_u^u$ by selection with $Q_{\text{min}}$
		\State  $E_c^{\mathrm{MRPT2}}[Q_u,P_m] \gets E^{\mathrm{MRPT2}}_c[Q_u^c, P_m] + E^{\mathrm{ENPT2}}_c[Q_u^u, P_m]$
		\EndFor
		\State For each subspace $V_i \subset Q_u$, extrapolate its contribution to $E_c^{\mathrm{MRPT2}}[Q_u,P_m]$
as function of $|E^{\mathrm{ENPT2}}_c[Q_u^u, P_m]|$  \Comment{$Q_{\text{min}}$ extrapolation}
        \State $E_c^{\mathrm{MRPT2}}[Q_t, P_m]\gets E_c^{\mathrm{ENPT2}}[P_s,P_m]+E_c^{\mathrm{MRPT2}}[Q_c, P_m]+E_c^{\mathrm{MRPT2}}[Q_u,P_m]$
		\State $E_{\text{tot}}[Q_t,P_m]\gets E^{(0)}[P_m]+E_c^{\mathrm{MRPT2}}[Q_t, P_m]$
		\EndFor
		\State  For each subspace $V_i$, extrapolate its contribution to $E_{\text{tot}}[Q_t, P_m]$ as function of $|E_c^{\mathrm{MRPT2}}[Q_t,P_m]|$ \Comment{$C_{\text{min}}$ extrapolation}
		\State \Return $E_{\text{tot}}[Q,P]$
	\end{algorithmic}
	\label{algo:iCI_MRPT2}
\end{algorithm}

\subsection{LCUT}\label{LCUT}
The $Q_u^c$ portion of the $Q_u$ space, as determined by the threshold $Q_{\text{min}}$ (say, $10^{-5}$),
contains typically $10^8$ CSFs. For iCI-MRMP2 and iCI-NEVPT2E, this is just
the dimension of $Q_u^c H_0^U Q_u^c$ in the coupled set of linear equations \eqref{LinearEQ1}, which is too costly.
Since iCI-MRMP2/NEVPT2E can be regarded as the coupled version of iCI-MRMP2D/NEVPT2,
the so-called `linear combination of uncoupled $\mathbf{t}$-amplitudes' (LCUT) adopted in
PASPT2\cite{PASPT2} can also be used here. Specifically,
 the first-order amplitudes $\mathbf{t}$ for the $Q_u^c$ space of iCI-MRMP2/NEVPT2E
 can be expressed as a linear combination of
the (uncoupled) first-order amplitudes $\mathbf{u}_l$ for the $Q_u^c$ space of iCI-MRMP2D/NEVPT2, viz.
\begin{align}
&(\mathbf{H}_{0,ll}^U-E_0\mathbf{I})\mathbf{u}_l=-\mathbf{V}_l^U,\label{u-equation}\\
&\mathbf{t} = \sum_l \mathbf{u}_l x_l, \label{eqn:LCUTApprox}
\end{align}
where $\mathbf{H}_{0,ll}^U$ is the matrix of $H_0^U$ over the XFs belonging to $Q_l^{(k)}$.
To determine the scalar expansion coefficients $\{x_l\}$, we invoke the Hylleraas functional
\begin{align}
\mathcal{H} &= \sum_{l,m} \langle\mathbf{u}_l|\mathbf{H}_{0,ll}^U-E^{(0)}\mathbf{I}|\mathbf{u}_m\rangle x_l x_m+2\sum_l\langle \mathbf{u}_l| \mathbf{V}^U_l\rangle x_l\\
&=\sum_{l,m; l\neq m} \langle\mathbf{u}_l|\mathbf{F}_{CO,lm} |\mathbf{u}_m\rangle x_lx_m + \sum_l \langle\mathbf{u}_l| \mathbf{V}^U_l\rangle (2x_l -x_l^2), \label{eqn:Hylleraas_func}
\end{align}
where use of Eq. \eqref{u-equation} to eliminate the diagonal term as well as the fact that only the $F_{CO}$ part of $H_0^U$ can couple XFs belonging to different subspaces
has been made to arrive at Eq. \eqref{eqn:Hylleraas_func}.
The minimization with respect to $x_l$ then gives rise to a small system of linear equations
\begin{equation}
	\sum_{m} \left(\langle\mathbf{u}_l|\mathbf{F}_{CO,lm} |\mathbf{u}_m\rangle-\delta_{lm}\langle\mathbf{u}_m|\mathbf{V}^U_m\rangle \right)  x_m=-\langle\mathbf{u}_l|\mathbf{V}^U_l\rangle. \label{HylleraasEq}
\end{equation}
The LCUT-based iCI-MRMP2/NEVPT2E (denoted as iCI-MRMP2*/NEVPT2E* hereafter) can be interpreted as an intermediate between
the uncoupled iCI-MRMP2D/NEVPT2 and the fully coupled iCI-MRMP2/NEVPT2E.
As can be seen from Table \ref{table:size_consistency_be4}, 
the LCUT approximation does not affect the size-consistency of iCI-NEVPT2, but tends to reduce marginally
the size-consistency errors of iCI-MRMP2.

\section{Implementation of iCI-MRPT2 }\label{sec:impl}
The iCI(2) approach is already available\cite{iCIPT2,iCIPT2New} in the MetaWave package\cite{MetaWave}.
The FIC-NEVPT2 approach for the contracted space $Q_c$ involves only contractions
between 1-RDMs/2-RDMs and molecular integrals and can thus be easily implemented.
This step is very cheap compared to the selection of the uncontracted space $Q_u$ and subsequent solution of the linear equations.
The latter is common to all variants of iCI-MRPT2 characterized by the zeroth-order Hamiltonians.
Following the unified implementation and MPI parallelization of wave function methods\cite{MetaWave,MPI_ICIPT2},
we tried to implement all iCI-MRPT2s in a unified manner, by making use of the existing infrastructure of MetaWave as much as possible.
In essence, different zeroth-order Hamiltonians can be resolved at compile time,
thanks to the type and tag systems provided by MetaWave\cite{MetaWave}.

\subsection{Selection of $Q_u$}\label{SecQuSelect}
As stated in Sec. \ref{SecSelection}, the selection of the 
uncontracted space $Q_u$ ($=V_A\oplus V_B\oplus V_C \oplus V_D \oplus V_F$) takes 
the ENPT1 amplitudes (determined by $H_0^U$) as the importance measure, which is itself 
computationally very expensive (compared to the iCI criteria in Eqs. \eqref{iCIS} and
\eqref{iCID}), needless to say the huge number of the oCFGs in the $Q_u$ space.
Therefore, further optimizations must be performed.
A particularly useful feature is that the $Q_l^{(k)}$ functions
(which are independent of the occupations of active orbitals) can be organized
in terms of the labels $l$ for the occupation patterns of the core and virtual orbitals.
Specifically, the $Q_{ab}^{(-2)}$ ($Q_{ij}^{(2)}$) functions can first be classified into
two types, one with $a=b$ ($i=j$) and the other with $a\ne b$ ($i\ne j$), and then ordered lexicographically.
There is only one class for the $Q_l^{(k)}$ functions belonging to $V_A$, $V_C$ and $V_D$.
This structure can be exploited to simplify the generation of the oCFGs $|I\rangle$ belonging to $Q_u$ and
identification of their connections to the interacting oCFGs $|J\rangle$ belonging to $P_m$, so as to
boost the computational performance by enabling dense BLAS operations through the reuse of the basic coupling coefficients (BCC)
($\langle I\mu | E_{pq} | J\nu\rangle$ and $\langle I\mu | e_{pq,rs} | J\nu\rangle$) and
pre‑ordered molecular integrals.

\subsubsection{Generation of $Q_u$ and identification of connections}
The naive approach to generate the $Q_u$ space and
determine its connection to the $P_m$ space is
to loop over all oCFGs $|J\rangle\in P_m$, generate all singly and doubly excited oCFGs $|I\rangle$, and record each pair $(I,J)$.
Sorting these pairs first by $I$ and then by $J$ would then identify all $|I\rangle \in Q_u$ that have nonzero matrix elements with
those $|J\rangle \in P_m$. However, this algorithm is too expensive.
To facilitate the discussion, we introduce single and double displacement operators
that can encode connection patterns by acting on the oCFGs.
Specifically, for a single excitation
we write $|I\rangle = D^p_q |J\rangle$ if $|I\rangle$ is obtained from $|J\rangle$ by moving an electron from the $q$-th to the $p$-th spatial orbital.
For a double excitation we write $|I\rangle = D^{pq}_{rs} |J\rangle$ if $|I\rangle$ is obtained from $|J\rangle$ by moving two electrons from the $r$-th and $s$-th to the $p$-th and $q$-th spatial orbitals, with the convention $p \le q$ and $r \le s$.
Such displacement operators commute with each another.
The concept can also be generalized to electron‑number non‑conserving situations (e.g., $D^p$ creates an electron in the $p$-th spatial orbital).
Note also that every oCFG $|I\rangle \in Q_u$ can be factorized as a product of an active and an inactive part,
$|I\rangle = |I_A\rangle \otimes |I_I\rangle$, and that
the displacement operator $D$ connecting $|I\rangle \in Q_u$ with $|J\rangle \in P_m$
can likewise be factorized as the product of an inactive and an active part, $D = D_I D_A$.
The inactive part (i.e., $Q_l^{(k)}$) need not be considered during the generation of connections, so that it is only necessary
to generate all active oCFGs $\{|I_A\rangle\}$ (determined by occupations of the active orbitals)
and determine their connections with the active oCFGs $\{|J_A\rangle\}$ in $P_m$. For instance, the
inactive replacement operators are $D^{a}$ ($D_i$) for the $V_A$ ($V_C$) subspace,  which means that
three-index active displacement operators $D^u_{tw}$ ($D^{uv}_t$)
are needed to generate the active oCFGs $\{|I_A\rangle\}$ and identify their connections with the active oCFGs $\{|J_A\rangle\}$ in $P_m$.
Storing all resulting connections in memory is yet impractical.
To resolve this, we make use of $n$-th order residues $\mathcal{R}_n$ composed of active oCFGs
generated by removing $n$ electrons from the active oCFGs $\{|J_A\rangle\}$ in $P_m$.
Each entry of $\mathcal{R}_n$ pairs a residue oCFG $|\mathbf{R}\rangle$ with a list of tuples $(J_A, \text{orb\_indices})$
that record which oCFG $|J_A\rangle \in P_m$ generates $|\mathbf{R}\rangle$ through which orbitals, thereby encoding the connection betweens $\mathcal{R}_n$ and $P_m$.
For instance, the active oCFGs in $V_A$ can be generated by applying all possible $D_t$
on each oCFG in $\mathcal{R}_{-2}$ that is formed by applying all possible $D^{uv}$ to each active oCFGs in $P_m$.
The connection between an active oCFG $|I_A\rangle$ and the active oCFGs $\{|J_A\rangle\}$ in $P_m$ can be identified
by first determining which oCFGs in $\mathcal{R}_{-2}$ are connected to $|I_A\rangle$,
and then taking the union of their associated lists of tuples.
This yields all $|J_A\rangle \in P_m$ that interact with $|I_A\rangle$.
The active oCFGs in $V_C$ and their connections to the active oCFGs in $P_m$ can be generated in the same way via residue $\mathcal{R}_2$.
The situation is somewhat simpler for the $V_B$, $V_D$ and $V_F$ subspaces, where two orbitals in the displacement operator are inactive,
so that the residues themselves ($\mathcal{R}_{-2}$, $\mathcal{R}_2$ and $\mathcal{R}_0$ for $V_B$,  $V_F$ and $V_D$, respectively)
directly encode the connections between the excited active oCFGs and the active oCFGs in $P_m$.
Here, the zeroth-order residue $\mathcal{R}_0$ for $V_D$ consists of all active oCFGs in $P_m$ and
those obtained by first removing an electron from the $u$-th orbital and then adding one to the $v$-th orbital, with $u \neq v$.
The full excited oCFGs together with their connections to the oCFGs in $P_m$ can finally be generated by attaching the corresponding inactive parts.

The above residue-based algorithm for generating the $Q_u$ space amounts to a refinement of that used in iCIPT2\cite{iCIPT2New,MetaWave,MPI_ICIPT2}.

\subsubsection{Basic coupling coefficients}
Having generated the oCFGs in $Q_u$ and their connections to the oCFGs in $P_m$,
it remains to see how to compute and reuse the BCCs. To this end,
we employ the tabular unitary group approach (TUGA)~\cite{iCIPT2,iCIPT2New},
which is particularly efficient for evaluating the BCCs between the CSFs of oCFGs with many open‑shell orbitals. Here,
the first step is to decompose the spin‑free second‑quantized Hamiltonian
\begin{equation}
H  =\sum_{p, q} h_{p q} E_{p q}+\frac{1}{2} \sum_{p, q, r, s}(p q \mid r s) e_{p q, r s}
\end{equation}
into a form that is consistent with the diagrams employed in the UGA\cite{Paldus1980} for the BCCs between pairs of CSFs, viz.
\begin{align}
		H= & H_1^0+H_2^0+H_1^1+H_2^1+H_2^2, \\
		H_1^0= & \sum_p h_{p p} E_{p p}, \\
		H_1^1= & \sum_{p<q} h_{p q} E_{p q}+\sum_{p>q} h_{p q} E_{p q}, \\
		H_2^0= & \frac{1}{2} \sum_p(p p | p p) E_{p p}\left(E_{p p}-1\right)
		        +\sum_{p<q}\left[(p p | q q) E_{p p} E_{q q}+(p q \mid q p) e_{p q, q p}\right], \\
		H_2^1= & \sum_{p \neq q}\left[(p p \mid p q)\left(E_{p p}-1\right) E_{p q}+(p q \mid q q) E_{p q}\left(E_{q q}-1\right)\right]\nonumber \\
		& +\sum_{p \neq q \neq r}\left[(p q \mid r r) E_{p q} E_{r r}+(p r \mid r q) e_{p r, r q}\right], \\
		H_2^2= & \sum_{p \leq r} \sum_{q \leq s}\left[2^{-\delta_{p r} \delta_{q s}}(p q \mid r s) e_{p q, r s}
+\left(1-\delta_{p r}\right)\left(1-\delta_{q s}\right)(p s \mid r q) e_{p s, r q}\right].
\end{align}
To reuse the BCCs, we note that the BCCs $\langle I\mu | E_{p q} | J\nu\rangle$ and $\langle I\mu | e_{p q, r s} | J\nu\rangle$
depend only on the relative occupation pattern of the oCFG pair $(I,J)$, but not on the specific orbital indices $\{p,q,r,s\}$.
Furthermore, orbitals that are either empty or doubly occupied in both $|I\rangle$ and $|J\rangle$
can be omitted because their segment values are just one in the Yamanouchi–Kotani phase convention.
As such, the BCCs can be reexpressed as $\langle I\mu | E_{\bar{p}\bar{q}} | J\nu\rangle$ and $\langle I\mu | e_{\bar{p}\bar{q}, \bar{r}\bar{s}} | J\nu\rangle$
in terms of the reduced orbital indices (ROI) $\{\bar{p}, \bar{q}, \bar{r}, \bar{s}\}$, which are obtained by deleting the common empty/doubly‑occupied orbitals in
$|I\rangle$ and $|J\rangle$.
The ROIs are one‑to‑one correspondent to the original orbital indices (see Table~\ref{tab:ROT_illustration} for an example).
In this way, the same BCC can be used for many integrals that share the same ROIs.
The ROIs together with a specific bra-ket occupation pattern constitute a reduced occupation table (ROT).
oCFG pairs of the same ROT share the same set of BCCs, thereby allowing reuse of the BCCs. This
can be enhanced even more dramatically for MRPT2,
because once an active oCFG $|I_A\rangle$ in $Q_u$ is connected to an active oCFG $|J_A\rangle$ in $P_m$,
any oCFG $|I\rangle$ in $Q_u$ with the same $|I_A\rangle$ but different inactive oCFGs $\{|I_I\rangle\}$
shares the same ROT for the pairs $(I,J)$, provided that
the inactive oCFGs $\{|I_I\rangle\}$ belong to the same class.
This is illustrated in Table~\ref{tab:ROT_illustration} for an example in the $V_F$ subspace.
Moreover, since the inactive part of a ROT is fixed,
the number of distinct ROTs for a given subspace is limited,
scaling as $O(N_o^{4-N_p-N_h}$, where $N_o$ is the maximal number of open‑shell orbitals in $P_m$ (see Table \ref{table:FOIS_MRPT2} for $N_p$ and $N_h$).
This gives $O(N_o^3)$ for $V_A$ and $V_C$, and $O(N_o^2)$ for $V_B$, $V_D$ and $V_F$.
Consequently, the distinct ROTs for $V_B$, $V_D$ and $V_F$ can be pre‑computed and stored in memory, in the
same spirit as the particle-hole correspondence-based graphical unitary group approach\cite{Wen1,Wen4,Wen6,Wen8},
where the external partial loops are pre‑tabulated, so that only the partial loops within the active space need to be searched.
The ROTs for $V_A$ and $V_C$ are still generated on the fly, as done in the implementation of ENPT2\cite{iCIPT2New,MetaWave,MPI_ICIPT2}.

\begin{table}[htbp]
	\small
	\centering
	\caption{Illustration of reduced orbital indices (ROI),
reduced occupation table (ROT), and how oCFGs in $V_F$ with the same active oCFG can share the same ROT.}
	\label{tab:ROT_illustration}
	\begin{threeparttable}
		\begin{tabular}{c|l|ccccc|cccc} \toprule
		&\multirow{2}{*}{OrbIndx}
		& \multicolumn{5}{c|}{Active Part}
		& \multicolumn{4}{c}{Inactive Part} \\ \cline{3-11}
		&& $0_a$ & $1_a$ & $2_a$ & $3_a$ & $4_a$
		& $5_v$ & $6_v$ & $7_v$ & $8_v$ \\\toprule
		\multirow{3}{*}{ROT1} &Bra1\tnote{a} & 1 & 0 & 0 & 2 & 0 & 1 & 1 & 0 & 0\\
		&Ket\tnote{b}  & 1 & 0 & 1 & 2 & 1 & 0 & 0 & 0 & 0\\
		&ROI      & $\bar{0}$ & x & $\bar{1}$ & x & $\bar{2}$ & $\bar{3}$ & $\bar{4}$ & x & x\\ \midrule
		\multirow{3}{*}{ROT2}&Bra2\tnote{a} & 1 & 0 & 0 & 2 & 0 & 0 & 0 & 1 & 1\\
		&Ket\tnote{b}  & 1 & 0 & 1 & 2 & 1 & 0 & 0 & 0 & 0\\
		&ROI      & $\bar{0}$ & x & $\bar{1}$ & x & $\bar{2}$ & x & x & $\bar{3}$ & $\bar{4}$ \\  \bottomrule
		\end{tabular}
		\begin{tablenotes}
			\item[a] oCFGs in $V_F$.
			\item[b] oCFGs in $P_m$.
		\end{tablenotes}
	\end{threeparttable}
\end{table}

\subsubsection{Evaluation of ENPT1 amplitudes}\label{OPtimizeBDF}
\begin{figure}[!ht]
	\includegraphics[width=0.6\textwidth]{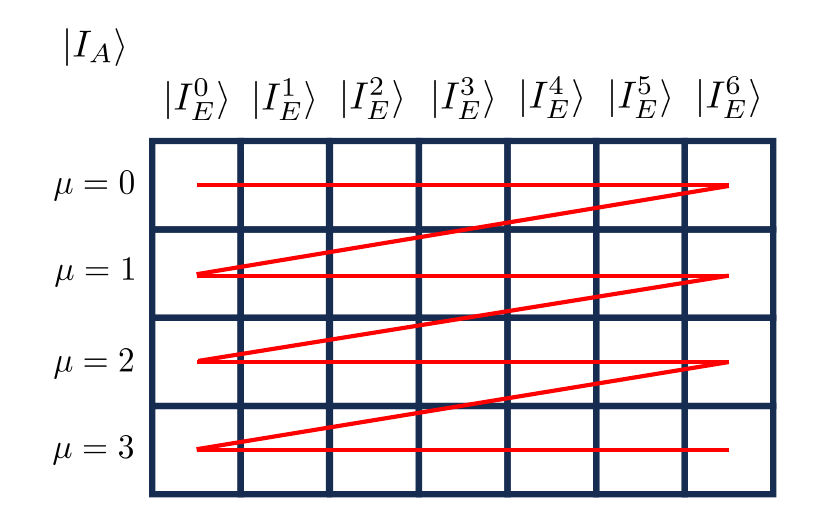}
	\caption{
	Storage layout of $\{\langle I\mu|H_0^U|\Psi^{(0)}[P_m]\rangle\}$ for oCFGs $|I\rangle\in Q_u$
	that share the same active part $|I_A\rangle$, illustrated for 7 distinct inactive oCFGs $|I_I^i\rangle$
($i\in[0,6]$) and 4 values of $\mu$.
	}
	\label{fig:numerator}
\end{figure}

The numerators of the ENPT1 amplitudes, $\langle I\mu|H_0^U|\Psi^{(0)}[P_m]\rangle$,
can be calculated by contractions between the molecular integrals and corresponding BCCs.
While there is no room of improvement for the $V_A$ and $V_C$ subspaces,
further optimization can be achieved for the $V_B$, $V_D$ and $V_F$ subspaces.
To show this, we introduce a list whose entries take the form
$\{ |I_0\mu\rangle, |J\nu\rangle, \text{BCC}, \text{orb\_indices}\}$,
where $|I_0\rangle$ is a representative oCFG chosen from the set of oCFGs $\{|I\rangle\}\in Q_u$
sharing the same active part $|I_A\rangle$ (and hence the same BCCs), $|J\nu\rangle$ is
used to retrieve the coefficient $C_{J\nu}^{(0)}$ of the zeroth‑order wave function $|\Psi_0\rangle[P_m]$,
and \text{orb\_indices} directs the fetching of the molecular integrals.
Since the two active orbital indices are fixed for a given $|I_A\rangle$, while the inactive orbital indices vary for different inactive oCFGs $|I_I\rangle$,
the molecular integral arrays can be classified by the active orbital indices, and each element of a specific array is labelled by an inactive oCFG.
By merging all such lists generated for a given pair $(I_0,J)$ over all $|J_A\rangle\in P_m$ connected to the given $|I_A\rangle$,
we obtain all information needed to compute the elements $\langle I\mu|H_0^U|\Psi^{(0)}[P_m]\rangle$ for the same $|I_A\rangle$ but different $|I_I\rangle$.
The corresponding array for $\{\langle I\mu|H_0^U|\Psi^{(0)}[P_m]\rangle\}$
is laid out as a row‑major matrix, with rows indexed by $\mu$ and columns corresponding to different inactive
oCFGs $|I_I\rangle$ (see Fig.~\ref{fig:numerator}).
The contribution of $|J\nu\rangle$ to the $\mu$-th CSF of all inactive oCFGs $\{|I_I\rangle\}$ is
obtained by calling a DAXPY operation that contracts the product $C_{J\nu}^{(0)} \times \text{BCC}$ with the array of molecular integrals
for a given pair of active orbitals. Such dense BLAS operations with contiguous access to the memory are much more
efficient than the sparse matrix-vector operations that would be involved in an oCFG-wise implementation.

To calculate the denominators of the ENPT1 amplitudes, we need the diagonal matrix elements of $H^U_0$, which can be expressed as\cite{iCIPT2}:
	\begin{align}
		\langle I \mu| H_0^U|I \mu\rangle
		&=   \frac{1}{2} \sum_p n_p^R\left[h_{p p}+\epsilon_p+\left(n_p^R-2\right) g_{p p}\right]\nonumber \\
		& + \sum_p \Delta_p^I\left[\epsilon_p+\left(n_p^R-1\right) g_{p p}\right]+\sum_{p \leq q} \Delta_p^I g_{p q} \Delta_q^I\nonumber  \\
		& + \sum_{p<q}(p q \mid q p)\langle I \mu| e_{p q, q p}|I \mu\rangle \delta\left(n_p^I, 1\right) \delta\left(n_q^I, 1\right),\label{eqn:dg_terms}\\
		\epsilon_p & = h_{p p}+\sum_r n_r^R g_{p r},\\
		g_{p r} & = (p p \mid r r)-\frac{1}{2}(p r \mid r p)=g_{r p}, \\
		\Delta_p^I & = n_p^I-n_p^R,
	\end{align}
where $\{n_p^R\}$ are the occupation numbers of a common reference configuration $|R\rangle$ (either Hartree--Fock or restricted open‑shell Hartree--Fock).
The first term in Eq.~\eqref{eqn:dg_terms} depends only on the reference configuration $|R\rangle$ and is therefore a uniform constant.
The second term is common to all CSFs of oCFG $|I\rangle$.
The only $\mu$-dependent contribution comes from the fourth term, where
the Kronecker deltas $\delta(n_p^I,1)\,\delta(n_q^I,1)$ restrict the sum to open‑shell orbitals.

The complexity of evaluating $\langle I\mu|H_0^U|I\mu\rangle$ increases in the following order:
the generalized Fock operator used in iCI‑MRMP2D, iCI‑MRMP2* and iCI‑MRMP2,
the Dyall Hamiltonian used in iCI‑NEVPT2D, iCI‑NEVPT2, iCI‑NEVPT2E* and iCI‑NEVPT2E,
and the full Hamiltonian used in iCI‑MRENPT2 (see Table \ref{iCI-MRPT2}).

For the generalized Fock operator,  Eq.~\eqref{eqn:dg_terms} is simplified to
	\begin{align}
		\langle I \mu| F|I \mu\rangle=\langle I \mu| F_{CD}|I \mu\rangle
		&=   \frac{1}{2} \sum_p n_p^R\left[h_{p p}+\epsilon_p \right]
		 + \sum_p \Delta_p^I\epsilon_p,
	\end{align}
where the summation over $p$ can further be decomposed into summations over the active and over the inactive orbitals,
so that
\begin{equation}
	\langle I \mu| F|I \mu\rangle=\langle I \mu| F_{CD}|I \mu\rangle = E^{(0)}_{I_A} + E^{(0)}_{I_I},
\end{equation}
where $|I\rangle = |I_A\rangle \otimes |I_I\rangle$, and
$E^{(0)}_{I_A}$ and $E^{(0)}_{I_I}$ are the contributions from $|I_A\rangle$
and $|I_I\rangle$, respectively.The latter is independent of $|I_A\rangle$ and can thus be pre‑computed.

The two-body term of the Dyall Hamiltonian $H^D$ is nonzero only when all orbital indices are active, so that
Eq.~\eqref{eqn:dg_terms} for $H^D$ can be written as
\begin{equation}
	\begin{aligned}
		\langle I \mu| H^D|I \mu\rangle
		=  & \frac{1}{2} \sum_p n_p^R\left[h_{p p}+\epsilon_p+\left(n_p^R-2\right) g_{p p}\right] \\
		& + \sum_u \Delta_u^I\left[\epsilon_u+\left(n_u^R-1\right) g_{u u}\right]+\sum_{u \leq v} \Delta_u^I g_{u v} \Delta_v^I \\
		& + \sum_{u<v}(u v \mid v u)\langle I \mu| e_{u v, v u}|I \mu\rangle \delta\left(n_u^I, 1\right) \delta\left(n_v^I, 1\right) \\
		& + \sum_i \Delta_i^I\epsilon_i + \sum_a \Delta_a^I\epsilon_a. \label{diagHD}\\
	\end{aligned}
\end{equation}
Here, the second and third lines involve only the active part $|I_A\rangle$ of oCFG $|I\rangle$. They are
independent of and dependent on the index $\mu$ of the CSFs, respectively.
The latter can be viewed as a function of $|I_A\mu\rangle$ and is identical for the $\mu$-th CSF of all oCFGs sharing the same $|I_A\rangle$.
The last line depends only on the inactive part $|I_I\rangle$ of $|I\rangle$ and can be evaluated in advance.
Consequently, $\langle I\mu|H^D|I\mu\rangle$ takes the following form
\begin{equation}
\langle I\mu|H^D|I\mu\rangle = E^{(0)}_{I_A\mu} + E^{(0)}_{I_I},
\end{equation}
where $E^{(0)}_{I_A\mu}$ collecting the first three lines in Eq. \eqref{diagHD}  is the contribution from $|I_A\mu\rangle$,
whereas $E^{(0)}_{I_I}$ is from $|I_I\rangle$.

The most complicated case is the full Hamiltonian used in iCI‑MRENPT2, which contains terms that couple the active and inactive orbital spaces.
Consequently, $\langle I\mu|H|I\mu\rangle$ can only be evaluated with Eq.~\eqref{eqn:dg_terms}.

In summary, by making proper organization of the inactive oCFGs independently of the active ones,
dense BLAS operations with contiguous access to the memory can be exploited for efficient evaluation of the numerators,
whereas the simplified forms for the diagonal elements of
the generalized Fock operator and Dyall Hamiltonian can be employed for efficient evaluation of the denominators
of the ENPT1 amplitudes. To estimate the effectiveness of these algorithmic optimizations,
we introduce average times $\Delta t_{Q_u}=T_{Q_u}/|Q_u|$ and $\Delta t_{\text{ENPT2}}=T_{\text{ENPT2}}/|P_s|$ for the contribution of
an XF to the $Q_u$-space ENPT1 amplitudes and to the ENPT2 energy in iCI(2), respectively. 
Assuming that the former employs the same algorithm as the latter,
and further ignoring the cost difference between $H_0^U$ and $H$, we would have $\Delta t_{Q_u}=\Delta t_{\text{ENPT2}}$.
Taking this as reference, the speedup furnished by the algorithmic optimizations for the ENPT1 amplitudes of
the $V_B$, $V_D$ and $V_F$ portions of $Q_u$ can be characterized by the factor   
$\Delta t_{\text{ENPT2}}/\Delta t_{Q_u}=(|Q_u|/|P_s|)\times (T_{\text{ENPT2}}/T_{Q_u})$.
This factor is meaningful particularly because the ENPT2 step in iCI(2) is much more expensive than the $P$-space selection itself,
so that the cost of the latter can be ignored. 
Take octacene as a showcase.
With the cc-pVTZ basis set\cite{Dunning1989}, the full CI space is (224e,1300o). The CAS(34e,34o)-iCI-NEVPT2 calculations
were performed with $C_{\text{min}} = \{10.0, 7.0, 5.0, 3.0, 1.5\}\times 10^{-5}$ and $Q_{\text{min}} = 5.0\times 10^{-6}$.
All timings were obtained on a single node equipped with one AMD EPYC 7H12 CPU (64 cores) and 1~TB of DDR4 memory.
As can be seen from Table~\ref{table:timing}, the speedup factor for iCI-NEVPT2 is approximately 9 for all values of $C_{\text{min}}$.

\begin{table}[htbp]
	\small
	\centering
	\caption{Wall times ($T_{\text{ENPT2}}$, $T_{Q_u}$ and $T_{\text{LE}}$ in seconds) and subspace dimensions for CAS(34e,34o)-iCI-NEVPT2/cc-pVTZ calculations of
octacene with $Q_{\text{min}} = 5.0\times 10^{-6}$.}
	\begin{threeparttable}
		\begin{tabular}{ccrrrrcrr} \toprule
			  \multirow{2}{*}{$C_{\text{min}}$}
			& \multirow{2}{*}{$|P_m|$}
			& \multicolumn{2}{c}{ENPT2 in iCI(2)}
			& \multicolumn{3}{c}{$Q_u$ selection}
			& \multicolumn{2}{c}{Linear Equations \eqref{LinearEQ1}} \\ \cline{3-9}
			&&
			\multicolumn{1}{c}{$|P_s|$} &
			\multicolumn{1}{c}{$T_{\text{ENPT2}}$} &
			\multicolumn{1}{c}{$|Q_u|$} &
			\multicolumn{1}{c}{$T_{Q_u}$} &
			\multicolumn{1}{c}{Speedup\tnote{a}} &
			\multicolumn{1}{c}{$|Q_u^c|$\tnote{b}} &
			\multicolumn{1}{c}{$T_{\text{LE}}$} \\ \toprule
			\multicolumn{1}{r}{$1.0\times 10^{-4}$}  & \multicolumn{1}{r}{110957} &
			$6.1\times 10^{9}$  & $1.9\times 10^{1}$ &
			$2.3\times 10^{13}$ & $7.0\times 10^{3}$ & 10.0 &
			$2.9\times 10^{8}$  & $8.6\times 10^{3}$ \\
			\multicolumn{1}{r}{$7.0\times 10^{-5}$}  & \multicolumn{1}{r}{188347} &
			$1.1\times 10^{10}$  & $2.7\times 10^{1}$ &
			$4.1\times 10^{13}$  & $1.3\times 10^{4}$ & 8.1 &
			$3.0\times 10^{8}$   & $7.7\times 10^{3}$ \\
			\multicolumn{1}{r}{$5.0\times 10^{-5}$}  & \multicolumn{1}{r}{317918} &
			$1.8\times 10^{10}$  & $4.4\times 10^{1}$ &
			$7.1\times 10^{13}$  & $1.9\times 10^{4}$ & 9.9 &
			$3.0\times 10^{8}$   & $9.2\times 10^{3}$ \\
			\multicolumn{1}{r}{$3.0\times 10^{-5}$}  & \multicolumn{1}{r}{701430} &
			$4.0\times 10^{10}$  & $8.9\times 10^{1}$ &
			$1.7\times 10^{14}$  & $5.0\times 10^{4}$ & 8.3 &
			$3.1\times 10^{8}$   & $8.2\times 10^{3}$ \\
			\multicolumn{1}{r}{$1.5\times 10^{-5}$}  & \multicolumn{1}{r}{1951431} &
			$1.1\times 10^{11}$  & $2.0\times 10^{2}$ &
			$5.0\times 10^{14}$  & $1.2\times 10^{5}$ & 8.9 &
			$3.1\times 10^{8}$   & $9.5\times 10^{3}$ \\  \bottomrule
		\end{tabular}
	\begin{tablenotes}
		\item[a] Defined as $(|Q_u|/|P_s|)\times (T_{\text{ENPT2}}/T_{Q_u})$ (see the text).
		\item[b] Size of the subspace selected with $Q_{\text{min}}$.
	\end{tablenotes}
	\end{threeparttable}\label{table:timing}
\end{table}

\subsection{Post-selection step}
Having selected the $Q_u$ space, it remains to solve the linear system of equations \eqref{LinearEQ1} for the amplitudes
of the $Q_u^c$ space,  and perform extrapolation with respect to $Q_{\text{min}}$. The former is
solved iteratively with the generalized minimal residual method (GMRES)~\cite{GMRES}.
The existing Hamiltonian construction code in MetaWave\cite{MetaWave} is used to build the matrix $\mathbf{H}_0$ of $H_0^U$.
Since $H_0^U$ is a one-body operator for iCI-MRMP2D, iCI-MRMP2* and iCI-MRMP2, double excitations are not involved.
The loop over double excitation can readily be removed at the compile time with no runtime cost,
thanks to the template metaprogramming techniques utilized in MetaWave\cite{MetaWave}.
Starting from an initial guess $\mathbf{t}_0$, both the residual $\mathbf{r}_i= (\mathbf{H}_0-E_0)\mathbf{t}_i+\mathbf{V}$ and the preconditioned residual $-(\mathbf{D}_{0}-E_0)^{-1} \mathbf{r}_i$ are added to the Krylov subspace of the previous iterations, after orthonormalization against the existing vectors.
Here $\mathbf{D}_{0}$ denotes the diagonal part of $\mathbf{H}_0$.
The iteration is restarted every $n$ (e.g., 8) steps, so that the Krylov subspace contains at most $2n$ vectors.
Convergence is reached when $\|(\mathbf{H}_0-E_0)\mathbf{t}+\mathbf{V}\|_2<\eta=10^{-9}$.
Having obtained the amplitudes of iCI-MRMP2D/NEVPT2, the linear system of equations \eqref{HylleraasEq}
can further be solved with GMRES for the amplitudes of iCI-MRMP2*/NEVPT2E* under the LCUT approximation, which
can further be iterated until convergence to obtain the amplitudes of iCI-MRMP2/NEVPT2E.
This version of GMRES has been implemented within MetaWave as a driver
class that takes as input a class encapsulating the matrix-vector product (MVP) operation.
All other linear algebra operations, which are independent of the MVP,
have been highly optimized in terms of memory usage and by leveraging high‑performance math libraries such as BLAS, LAPACK, and MKL.
The iCI‑MRPT2 family of methods reuses this part of the MetaWave infrastructure, thereby achieving high performance while minimizing development effort.

For a given $C_{\text{min}}$, the selection of $Q_u$ with different values of $Q_{\text{min}}$  can be accomplished with a single
pass using the smallest $Q_{\text{min}}$, because the selected space $Q_u^c$ ($=Q_u-Q_u^u$)
 for a larger $Q_{\text{min}}$ is always a subset of that for a smaller $Q_{\text{min}}$.
Note in passing that the size of $Q_u^c$ with the same $Q_{\mathrm{min}}$ is rather insensitive to
a range of $C_{\text{min}}$,
so that the cost for solving the system of linear equations \eqref{LinearEQ1}
remains nearly constant and becomes less significant for smaller $C_{\text{min}}$, see Table \ref{table:timing}.
The second-order correlation energies $E^{\mathrm{MRPT2}}_c[Q_u^c,P_m](Q_{\mathrm{min}})$ and $E^{\mathrm{ENPT2}}_c[Q_u^u,P_m](Q_{\mathrm{min}})$
can be decomposed into contributions from individual subspaces $V_X$ ($X=A, B, C, D, V$), and 
$E^{\mathrm{MRPT2}}[V_X,P_m](Q_{\mathrm{min}})
= E^{\mathrm{MRPT2}}_c[V_X^c,P_m](Q_{\mathrm{min}})+ E^{\mathrm{ENPT2}}_c[V_X^u,P_m](Q_{\mathrm{min}})$ for a given $C_{\mathrm{min}}$
can be extrapolated to the zero limit of $|E^{\mathrm{ENPT2}}_c[V_X^u,P_m]|(Q_{\mathrm{min}})$, leading to $E^{\mathrm{MRPT2}}[V_X,P_m](Q_{\mathrm{min}}=0)$.
In this work, we employ a weighted linear extrapolation with weights $1/|E^{\mathrm{ENPT2}}_c[V_X^u,P_m](Q_{\mathrm{min}})|$,
so that data points with smaller $|E^{\mathrm{ENPT2}}_c[V_X^u,P_m](Q_{\mathrm{min}})|$ carry more weights.
The total MRPT2 energy $E_{\mathrm{tot}}[Q_t,P_m](Q_{\mathrm{min}}=0)$ can then be extrapolated to the zero limit of $|E^{\mathrm{ENPT2}}_c[P_s,P_m]|$,
so as to approach the CAS-based iCI-MRPT2 energy $E_{\mathrm{tot}}[Q,P]$.

\clearpage		
\newpage

\section{RESULTS AND DISCUSSION}\label{sec:pilot_app}
The iCISCF approach\cite{iCISCF} with $C_{\text{min}}=5\times 10^{-5}$ was used to optimize the orbitals,
which were canonicalized upon convergence. Such calculations were performed with the BDF package\cite{BDF1,BDF2020,BDF2026},
which provides molecular integrals for subsequent iCI-MRPT2 calculations with the MetaWave package\cite{MetaWave,MetaWaveURL}.
Unless otherwise specified, the iCI‑MRPT2 energies were extrapolated using the sequence $C_{\text{min}} = \{5,3,1.5,0.9,0.7\}\times10^{-5}$.
For each $C_{\text{min}}$, the corresponding $Q_{\text{min}}$ values were $\{30,15,9,7,5\}\times10^{-6}$.
For some systems considered here, iCI-MRMP2D, iCI-MRMP2 and iCI-MRMP2* suffer from the intruder-state problem, contrary to 
iCI‑NEVPT2, iCI‑NEVPT2E*, iCI‑NEVPT2E, iCI-NEVPT2D and iCI-MRENPT2. A direct consequence is that the GMRES solver for the linear equations does not converge in the former, 
whereas it can converge typically within 10 iterations in the latter. For this reason, iCI-MRMP2D, iCI-MRMP2 and iCI-MRMP2*
were applied only to polyacenes. 

\subsection{Extrapolation}
\begin{figure}[!ht]
	\includegraphics[width=1.0\textwidth]{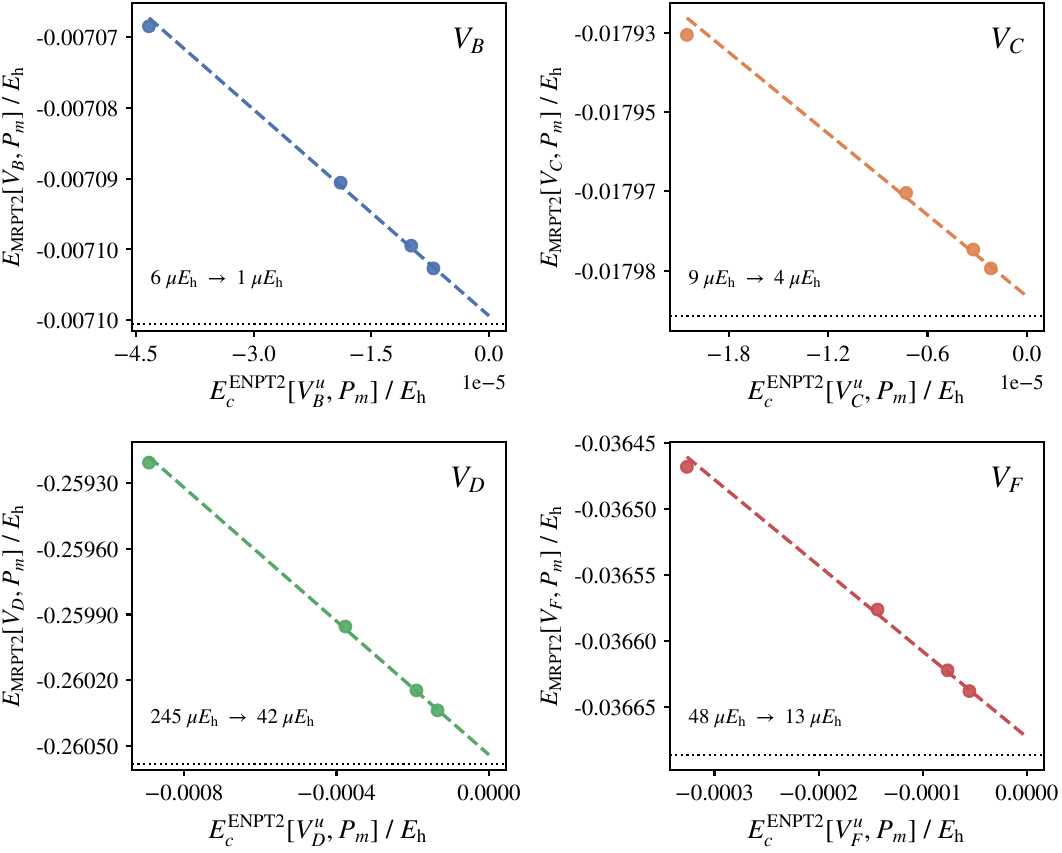}
	\caption{$Q_{\mathrm{min}}$ extrapolation at $C_{\mathrm{min}}=0.0$. For each subspace $V_X\subset Q_u$,
	$E^{\mathrm{NEVPT2}}[V_X,P_m]=E^{\mathrm{NEVPT2}}_c[V_X^c,P_m]+E^{\mathrm{ENPT2}}_c[V_X^u,P_m]$ is plotted against $E^{\mathrm{ENPT2}}_c[V_X^u,P_m]$
for two non-interacting benzenes with (12e, 12o) and cc-pVDZ. $Q_{\mathrm{min}}=\{30,15,9,7\}\times 10^{-6}$.
Deviations of the results with $Q_{\mathrm{min}}=7\times 10^{-6}$ and extrapolation from the CAS-based value (horizontal dotted line)
are labeled in each subfigure. 
		}\label{fig:benzene_extra_qmin}
\end{figure}

\begin{figure}[!ht]
	\includegraphics[width=1.0\textwidth]{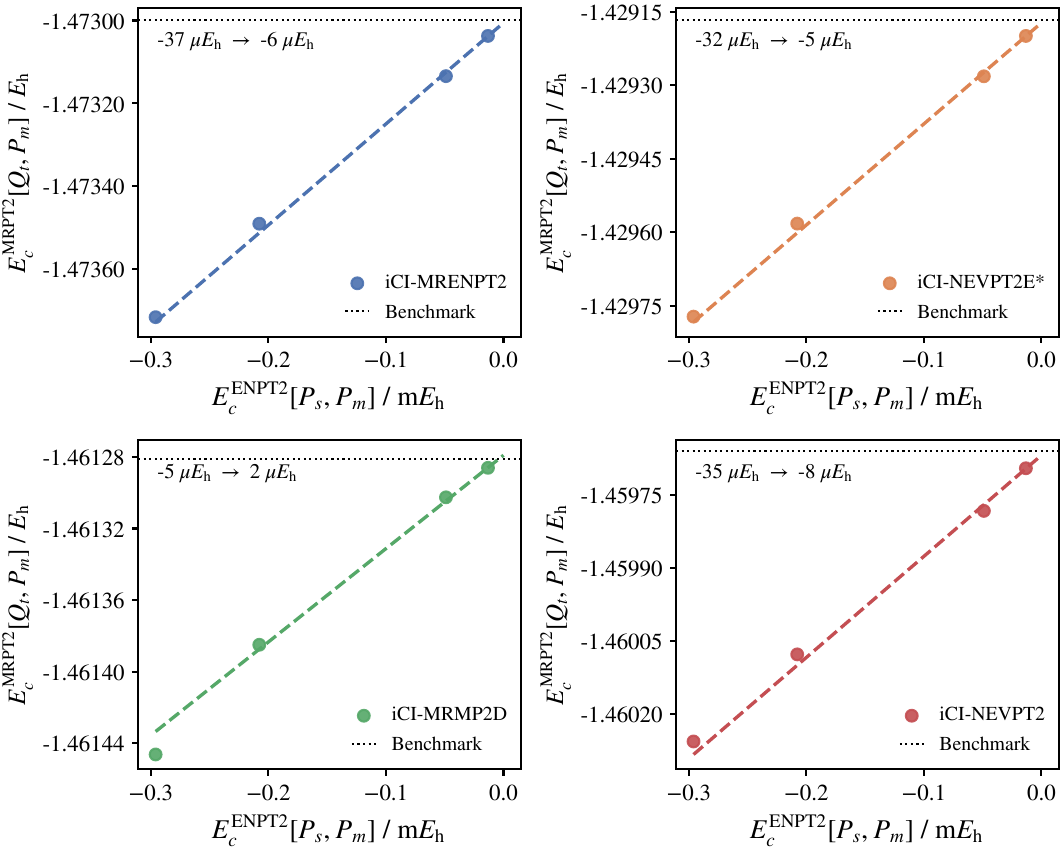}
	\caption{$C_{\text{min}}$ extrapolation of $Q_{\text{min}}$-extrapolated $E^{\text{MRPT2}}_c[Q_t,P_m]$ with respect to $E^{\text{ENPT2}}_c[P_s,P_m]$
for two non-interacting benzenes with CAS(12e,12o) and cc-pVDZ. 
	The $C_{\text{min}}$ values $\{5.0,3.0,1.0,0.5\}\times 10^{-4}$ are used for weighted linear extrapolation with weights $1/|E^{\text{ENPT2}}_c[P_s,P_m]|$.
Deviations of the results with $C_{\mathrm{min}}=5.0\times 10^{-5}$ and extrapolation from the CAS-based value (horizontal dotted line)
are labeled in each subfigure. 
	} \label{fig:benzene_extra_cmin}
\end{figure}

It is well established\cite{sCIextrapolation} in the context of sCI that, once the $E_{\mathrm{tot}}$ vs $|E_c^{(2)}|$ plot has reached the linear regime,
near-FCI result can be obtained by extrapolation. This should also hold for PAS-based against CAS-based iCI-MRPT2, albeit with two extrapolations:
a $Q_{\text{min}}$ extrapolation followed by a $C_{\text{min}}$ extrapolation. To show this, we take two non-interacting 
benzene molecules, described by CAS(12e,12o) and the cc-pVDZ basis set. The geometry of each benzene molecule was taken from Ref.~\citenum{benzene}.
Fig. \ref{fig:benzene_extra_qmin} reveals that the $Q_{\text{min}}$-extrapolation does approach the zero $Q_{\text{min}}$ limit for each subspace 
$Q_l^{(k)}$ of $Q_u$, while Fig. \ref{fig:benzene_extra_cmin} further shows that the subsequent $C_{\text{min}}$-extrapolation
can approach the CAS-based iCI-MRPT2 energies. 

A more challenging system is \ce{Cr2}, which requires simultaneous treatment
of relativistic and strong correlation effects. It has been investigated by a variety of computational methods, as summarized in Table~\ref{table:cr2_benchmark}.
The minimal active space (12e,12o), composed of the 3d and 4s atomic orbitals, is known to be insufficient, generally leading to overestimated dissociation energies~\cite{TDNEVPT2_1,TDNEVPT2_2,DMRG_sc_NEVPT2}. Therefore, the present iCI‑NEVPT2, iCI‑NEVPT2E* and iCI‑NEVPT2E calculations
took (12e,22o) as the active space, which includes the 3d, 4s, and 4d atomic orbitals.
All electrons were correlated in conjunction with the spin‑free exact two‑component (X2C) Hamiltonian~\cite{X2CSOC1,X2CSOC2} and 
the cc‑pwCVXZ‑DK basis sets  (X = T, Q, 5). 

\begin{figure}[!ht]
	\includegraphics[width=0.9\textwidth]{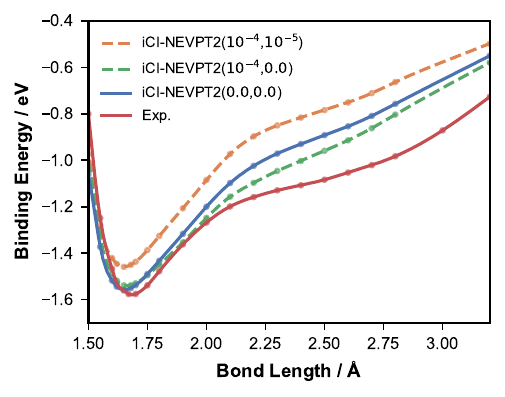}
	\caption{Ground-state potential energy curves of \ce{Cr2} by  extrapolated CAS(12e,22o)-iCI-NEVPT2($C_{\text{min}}$, $Q_{\text{min}}$)/cc-pwCV5Z-DK.
	'0.0' refers to the extrapolated result.
	} \label{fig:cr2_extra}
\end{figure}

\begin{figure}[!ht]
	\includegraphics[width=0.9\textwidth]{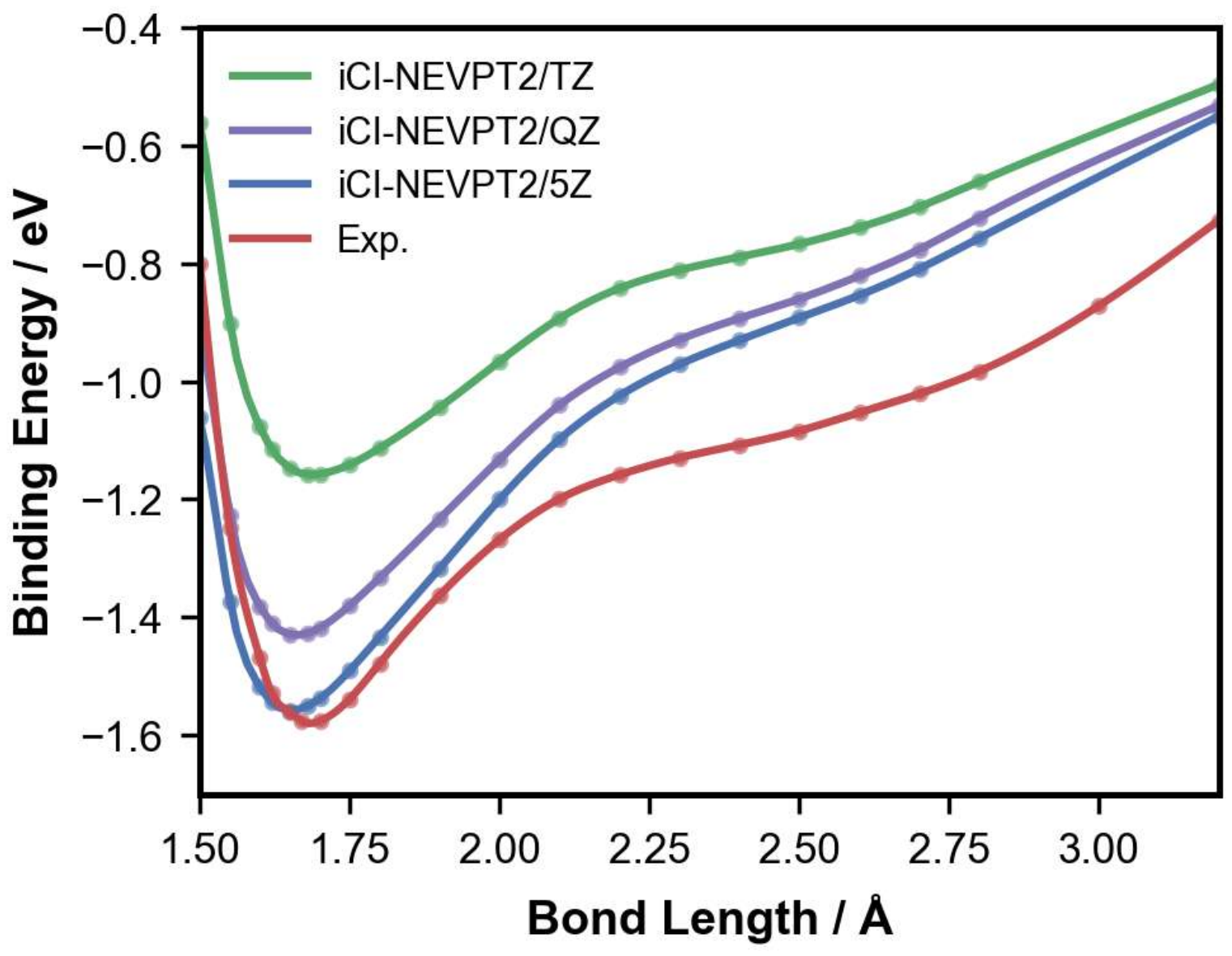}
	\caption{Ground-state potential energy curves of \ce{Cr2} by extrapolated CAS(12e,22o)-iCI-NEVPT2/cc-pwCVXZ-DK (X = T, Q, 5).}
	\label{fig:cr2_basis}
\end{figure}

The iCI-NEVPT2($C_{\text{min}}, Q_{\text{min}}$) potential energy curve of \ce{Cr2} 
with thresholds $ (10^{-4}, 10^{-5})$ is first compared in Fig.~\ref{fig:cr2_extra}
with the $Q_{\text{min}}$-extrapolated one and also the fully extrapolated one. It can be seen that, although  
the three curves are all smooth,  the $Q_{\text{min}}$-extrapolation has a pronounced effect on the curve shape,
which is further refined by the subsequent $C_{\text{min}}$-extrapolation, especially at distances longer than 2.0~\AA. 
It can also be seen from Fig.~\ref{fig:cr2_basis} that the fully extrapolated iCI-NEVPT2 
energy curve approaches steadily the experimental one as the basis set is enlarged (see Tables S1 and S2 in the Supporting Information).
However, the iCI-NEVPT2/5Z curve still differs significantly from the experiment one at distances longer than 2.0~\AA. 
Nevertheless, the equilibrium bond lengths, harmonic vibrational frequencies and dissociation energies 
by iCI‑NEVPT2/5Z (see Table~\ref{table:cr2_benchmark})
agree well with the theoretical best estimates (TBEs)~\cite{Cr2_TBE}, with the deviations being
only $0.03$\,\AA, $4$\,cm$^{-1}$ and $0.02$\,eV. 

\begin{figure}[!ht]
	\includegraphics[width=0.9\textwidth]{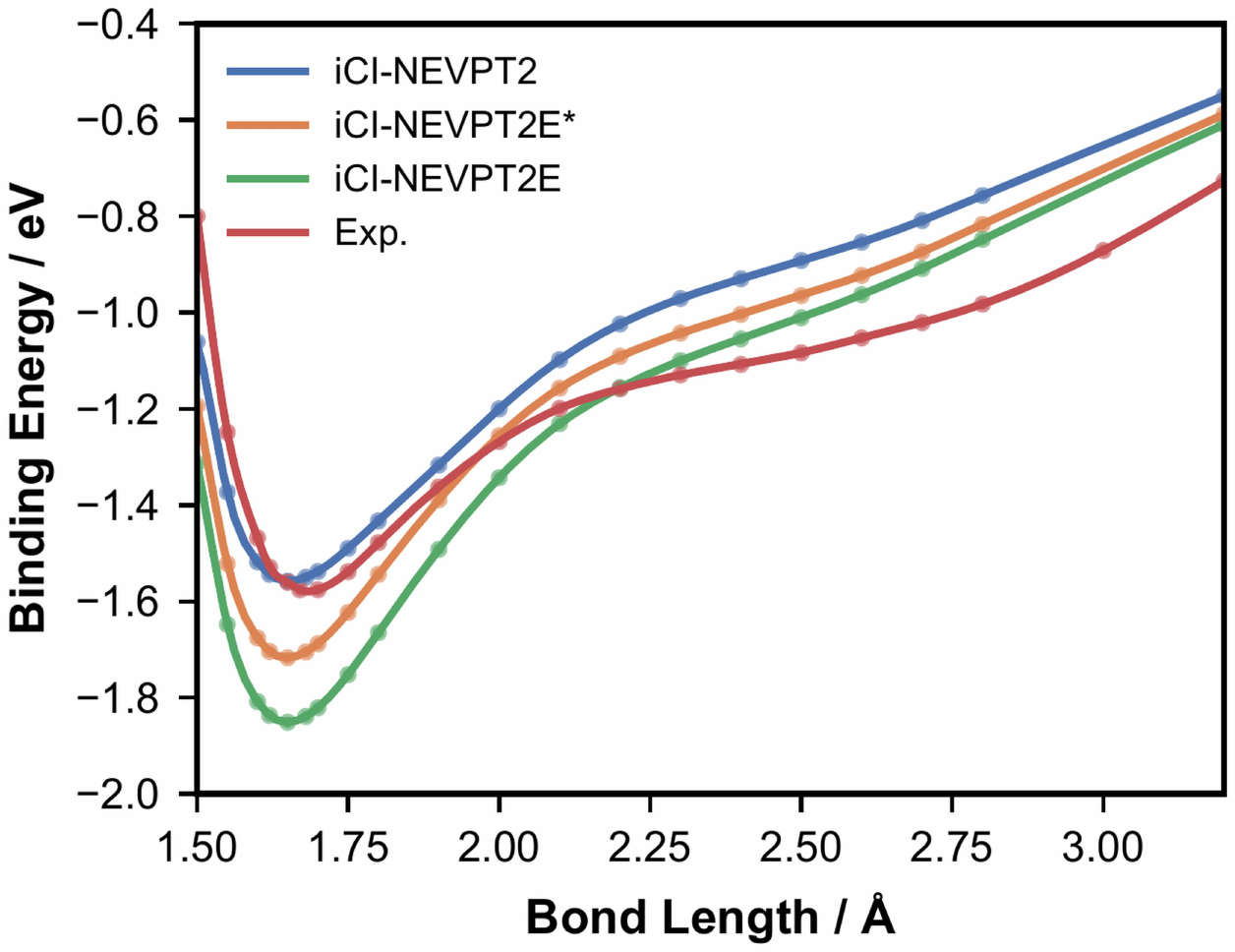}
	\caption{Ground-state potential energy curves of \ce{Cr2} by  extrapolated iCI-MRPT2 with CAS(12e,22o) and cc-pwCV5Z-DK.}
	\label{fig:cr2_method}
\end{figure}

\begin{figure}[!ht]
	\includegraphics[width=0.9\textwidth]{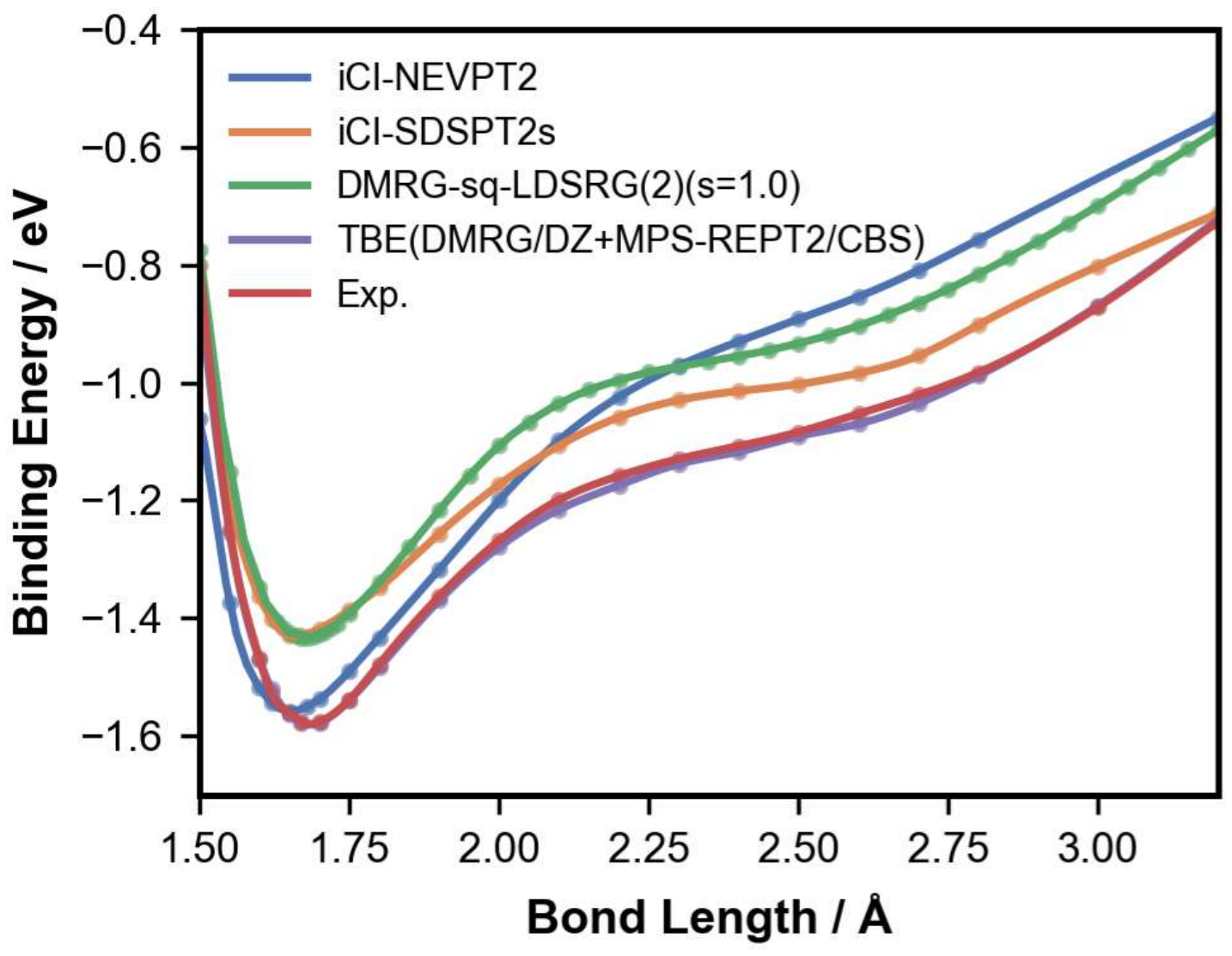}
	\caption{Ground-state potential energy curves of \ce{Cr2} by different methods and experiment (cf. Table \ref{table:cr2_benchmark}). 
	}\label{fig:cr2_cmpr_other_methods}
\end{figure}

Comparing the iCI-NEVPT2E and iCI-NEVPT2 results in Table~\ref{table:spectro_ici_mrpt2} and Fig.~\ref{fig:cr2_method}
reveals that the inclusion of the class-off-diagonal Fock operator $F_{CO}$ \eqref{FcoOper} in the former has little influence on $R_e$, 
but yields larger $\omega_e$ and $D_e$ across the three basis sets.
The LUCT approximation (iCI-NEVPT2E*) to iCI-NEVPT2E tends to reduce the $D_e$ of the latter by 0.15 eV, but does not
affect both $R_e$ and $\omega_e$. Noticeably,  
iCI‑NEVPT2E performs the worst among the three variants of iCI-NEVPT2, albeit most expensive. However, this should be viewed with caution. As 
shown in Table S3 in the Supporting Information, when a Morse fit and basis set extrapolation are employed, 
the iCI‑NEVPT2E results ($R_e = 1.682$\,\AA, $\omega_e = 481.4$\,cm$^{-1}$, $D_e = 1.557$\,eV) 
are in excellent agreement with the experimental values ($1.68$\,\AA, $479$\,cm$^{-1}$, $1.56$\,eV). 
This indicates that the fitting procedure for deriving molecular spectroscopic constants can play a significant role. 
Compared to other theoretical methods (Table~\ref{table:cr2_benchmark} and Fig.~\ref{fig:cr2_cmpr_other_methods}), 
iCI‑NEVPT2 tends to predict a slightly shorter bond length.
Yet, with respect to the TBEs, iCI‑NEVPT2/5Z is the only method that well reproduces both $\omega_e$ and $D_e$.

P

\begin{table}[htbp]
\small
\centering
\caption{Equilibrium bond lengths $R_e$ (in \AA), harmonic vibrational frequencies $\omega_e$ (in cm$^{-1}$) 
and dissociation energies $D_e$ (in eV) of \ce{Cr2} calculated by extrapolated iCI-NEVPT2, iCI-NEVPT2E* and iCI-NEVPT2E 
with active space (12e,22o).}
\begin{threeparttable}
\begin{tabular}{ccSSS} \toprule
method
& basis set
& \multicolumn{1}{c}{$R_e$}
& \multicolumn{1}{c}{$\omega_e$\tnote{a}}
& \multicolumn{1}{c}{$D_e$} \\\toprule
iCI-NEVPT2
& TZ & 1.687 & 380.1 & 1.158 \\
& QZ & 1.660 & 447.8 & 1.430 \\
& 5Z & 1.651 & 483.4 & 1.558 \\
& CBS(QZ/5Z) & 1.645 & 519.4 & 1.668 \\ \midrule
iCI-NEVPT2E*
& TZ & 1.676 & 446.2 & 1.302 \\
& QZ & 1.655 & 515.9 & 1.582 \\
& 5Z & 1.649 & 541.0 & 1.717 \\
& CBS(QZ/5Z) & 1.644 & 559.1 & 1.834 \\ \midrule
iCI-NEVPT2E
& TZ & 1.675 & 476.7 & 1.430 \\
& QZ & 1.656 & 549.3 & 1.715 \\
& 5Z & 1.649 & 542.8 & 1.851 \\
& CBS(QZ/5Z) & 1.642 & 564.2 & 1.970 \\ \midrule
\end{tabular}
\begin{tablenotes}
	\item[a] Spline fitting by the orca\_pesfit program.
\end{tablenotes}
\end{threeparttable} \label{table:spectro_ici_mrpt2}
\end{table}

\begin{table}[htbp]
	\small
	\centering
	\caption{Comparison of different methods for the equilibrium bond length $R_e$ (in \AA), harmonic vibrational frequency $\omega_e$ (in cm$^{-1}$) 
and dissociation energy$D_e$ (in eV) of \ce{Cr2}.}
	\begin{threeparttable}
		\begin{tabular}{lclScS} \toprule
			\multicolumn{1}{c}{method}
			& active space
			& \multicolumn{1}{c}{basis set}
			& \multicolumn{1}{c}{$R_e$}
			& \multicolumn{1}{c}{$\omega_e$}
			& \multicolumn{1}{c}{$D_e$} \\\toprule
			iCI-NEVPT2               & (12e,22o)	& cc-pwCV5Z-DK & 1.651 & 483.4 & 1.558 \\
			                         &              & CBS(QZ/5Z)   & 1.645 & 519.4 & 1.668 \\
			    SDSPT2s\tnote{a}     & (12e,22o)	& cc-pwCV5Z-DK & 1.659 & 462 & 1.45  \\
			DMRG-DSRG-PT3/$s=0.5$\tnote{b}    & (12e,22o)	& CBS(QZ/5Z)   & 1.655 & 579 & 1.44  \\
			DMRG-sqLDSRG(2)/$s=1.0$\tnote{c}  & (12e,22o)	& cc-pwCV5Z-DK & 1.676 & 473 & 1.43  \\
			                                  &             & CBS(QZ/5Z)   & 1.672 & 479 & 1.45  \\
			DMRG-sc-NEVPT2\tnote{d}           & (12e,22o)	& CBS(QZ/5Z)   & 1.656 & 470 & 1.45  \\
			DMRG-ec-MRCISD+Q\tnote{e}         & (12e,42o)	& ANO-RCC-VQZP & 1.71  & 477 & 1.62  \\
			MRAQCC\tnote{f}                   & (12e,12o)	& CBS(TZ/QZ)   & 1.685 & 459 & 1.36  \\
			TBE\tnote{g}                      &             & CBS(QZ/5Z)   & 1.685 & 495 & 1.58  \\ \midrule
			Expt.                    &              &              & 1.68\tnote{h} & 479\tnote{h} & \multicolumn{1}{c}{$1.472\pm0.052$\tnote{h}}\\
			                         &              &              &&&
			                         \multicolumn{1}{c}{$1.56\pm 0.06$\tnote{i}}\\\bottomrule
		\end{tabular}
		\begin{tablenotes}
			\item[a] Ref. \citenum{{SDSPT2s}}.
			\item[b] Ref. \citenum{DSRG2025}. Density matrix renormalization group 
(DMRG)-based driven similarity renormalization group third-order perturbation theory (DMRG $M=1000$).
			\item[c] Ref. \citenum{DSRG2025}. DMRG-based sequential version of linear two-body approximation of DSRG (DMRG $M=1000$).
			\item[d] Ref. \citenum{DMRG_sc_NEVPT2}. DMRG-based strongly contracted NEVPT2 (DMRG $M=800$).
			\item[e] Ref. \citenum{ecmrciDMRG}. DMRG-based externally contracted multireference configuration interaction with singles and doubles and Pople correction 
(DMRG $M=1000$).
			\item[f] Ref. \citenum{MRAQCC_Cr2}. Multireference averaged quadratic coupled cluster.
			\item[g] Ref. \citenum{Cr2_TBE}.
			\item[h] Ref. \citenum{Cr2Exp1}. Theoretical best estimate. 
			\item[i] Ref. \citenum{Cr2Exp2}. The zero-point energy of 0.03 eV is included.
		\end{tablenotes}
\end{threeparttable} \label{table:cr2_benchmark}
\end{table}

\clearpage
\newpage

\subsection{Polyacenes}

Polyacenes are another classic test ground for quantum chemical methods \cite{DSRG2025,ecmrciDMRG,PolyAceneGASGeom,PolyAceneGASGeomCCSDT,iCISCF,ASCIMRPT2,ASCI_DSRG,ASCISCF_OptGeom,ICEFRNEPT2,ACI2018}, 
as summarized in Table \ref{table:polyacene}.
The focus here is the adiabatic singlet--triplet gaps of $k$-acenes with $k\in[2,10]$, with 
the geometries taken from Ref.~\citenum{ASCISCF_OptGeom}.
Before a detailed discussion, it is worth mentioning that iCI-MRMP2 differs very little from iCI-MRMP2D,   
not only in the energy gaps but also in the total energies (ca. 1.0 m$E_{\text{h}}$), see  
Table S5 in the Supporting Information. 
A similar trend is also observed between iCI-NEVPT2E and iCI-NEVPT2.  
However, this is not a general behavior but is due to symmetry reasons. Specifically,
the active $\pi$-tipe orbitals $\{|u\rangle\}$ and core $\sigma$-type orbitals $\{|i\rangle\}$ belong to different 
irreducible representations of the $D_{2h}$ point group, so that $F_{\mathrm{co}}=F_{\mathrm{oc}}^\dag=0$, 
and meanwhile the $Q_{,i}^{(1)}$ subspace involving core-to-active single excitations is empty.
It can then be deduced from Table~\ref{table:coupling_PQ} that 
only $Q_{a,}^{(-1)}$ and $Q_{ab,}^{(-2)}$ are mutually coupled, whereas 
both $Q_{,ij}^{(2)}$ and $Q_{a,i}^{(0)}$ remain uncoupled. It happens that the former two
subspaces have opposite, cancelling contributions. 
This explains the observed trends. 
Therefore, we focus on iCI-MRMP2D and iCI-NEVPT2 in the remainder of this section.

\begin{figure}[!ht]
	\includegraphics[width=0.9\textwidth]{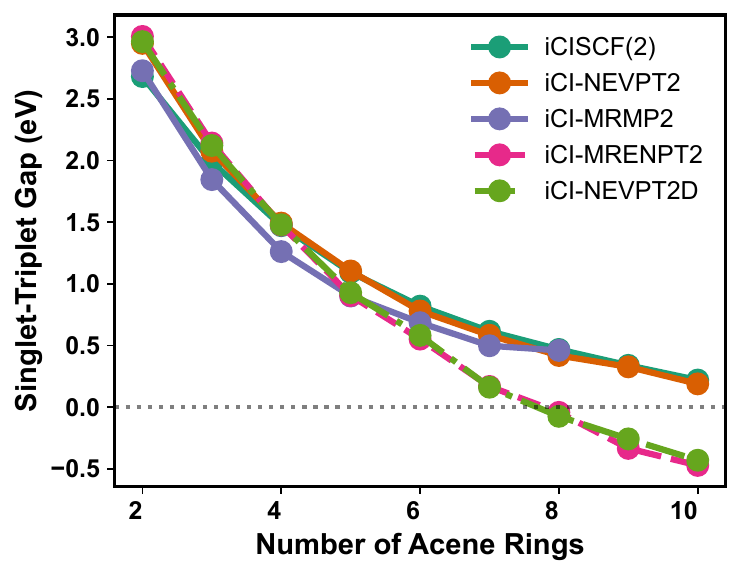}
	\caption{Singlet-triplet gaps of k-acenes ($k\in[2,10]$) by iCISCF(2), iCI-NEVPT2, iCI-MRMP2, iCI-MRENPT2 and iCI-NEVPT2D with cc-pVTZ and active spaces [($4k+2$)e, ($4k+2$)o].}
	\label{fig:method_comparison_ici}
\end{figure}

As shown in Fig.~\ref{fig:method_comparison_ici}, the inclusion of dynamic correlation tends to
reduce the iCISCF(2) adiabatic singlet--triplet gaps, and this reduction becomes smaller as the system size increases. 
This tendency is also observed in previous studies\cite{DSRG2025,ecmrciDMRG,PolyAceneGASGeom,PolyAceneGASGeomCCSDT,ICEFRNEPT2,ASCIMRPT2}.
Both iCI-MRENPT2 and iCI-NEVPT2D predict overly negative gaps for larger acenes by
overstabilizing the triplet states, which is inconsistent with other methods listed in Table \ref{table:polyacene}.
Since they are neither orbital-invariant nor size-consistent, they should not be recommended.
It says simply that a large active space does not necessarily mean that a crude zeroth-order Hamiltonian 
can be adopted for outer-space dynamic correlation.

When comparing iCI-MRMP2D and iCI-NEVPT2, the gaps by the former are systematically smaller than those by the latter, 
mirroring the relation between the ASCI-CASPT2D\cite{ASCIMRPT2} and ASCI-SC-NEVPT2\cite{ICEFRNEPT2} gaps.
This should be attributed to the use of different zeroth-order Hamiltonians.
On the other hand, the trend  that the iCI-NEVPT2 (iCI-MRMP2D) gaps are larger by ca. 0.10 eV than those by ASCI-SC-NEVPT2/ICE-FR-NEVPT2 (ASCI-CASPT2D)
should be ascribed to the use of different contractions (partial vs FIC). It can also be seen that  
the iCI-MRMP2D gaps are similar to those by DMRG-ec-MRCISD+Q\cite{{ecmrciDMRG}} (up to $k=7$), while 
the iCI-NEVPT2 gaps resemble those by CCSD(T)\cite{PolyAceneGASGeomCCSDT} and SS-DMRG-DSRG-PT3\cite{DSRG2025}.
However, given the sizeable extrapolation uncertainties for larger acenes ($k\geq 7$, see Table S8 and S9 in the Supporting Information), 
the comparison remains to be sharpened by performing calculations with smaller $C_{\text{min}}$ and $Q_{\text{min}}$.
To achieve this, MPI parallelization of the iCI-MRPT2 workflows should be implemented to enable the use of greater computational resources.
Furthermore, advanced extrapolation strategies should also be developed to reduce extrapolation uncertainties.
Research along these directions is actively pursued at our laboratory.

\begin{figure}[!ht]
	\includegraphics[width=0.9\textwidth]{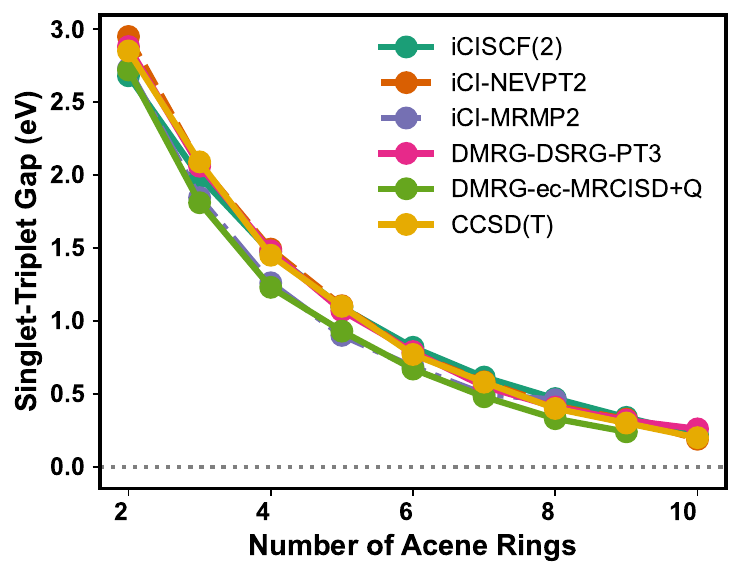}
	\caption{Singlet-triplet gaps of k-acenes ($k\in[2,10]$) by various methods.
	For more details, see Table \ref{table:polyacene}.
	}
	\label{fig:method_comparison_others}
\end{figure}

\begin{sidewaystable}
	\tiny
	\centering
	\caption{Adiabatic Singlet-Triplet Gaps (in eV) of k-acenes ($k\in[2,10]$)}
	\begin{threeparttable}
	\begin{tabular}{cSSSSSSSSS} \toprule
		$k$
		&\multicolumn{1}{c}{SS-DMRG-DSRG-PT3\tnote{a}}
		&\multicolumn{1}{c}{DMRG-ec-MRCISD+Q\tnote{b}}
		&\multicolumn{1}{c}{DMRG-GAS-PDFT\tnote{c}}
		&\multicolumn{1}{c}{CCSD(T)\tnote{d}}
		&\multicolumn{1}{c}{ICE-FR-NEVPT2\tnote{e}}
		&\multicolumn{1}{c}{ASCI-CASPT2D\tnote{f}}
		&\multicolumn{1}{c}{ASCI-SC-NEVPT2\tnote{f}}
		&\multicolumn{1}{c}{iCI-MRMP2D\tnote{g}}
		&\multicolumn{1}{c}{iCI-NEVPT2\tnote{g}} \\ \toprule
		2 &2.88 &2.72                 &2.81 &2.85 &\multicolumn{1}{c}{-}&\multicolumn{1}{c}{-}&\multicolumn{1}{c}{-}&2.73 &2.95 \\
		3 &2.06 &1.81                 &1.87 &2.09 &2.01                 &1.71                 &2.01                 &1.85 &2.08 \\
		4 &1.48 &1.23                 &1.25 &1.45 &1.38                 &1.16                 &1.39                 &1.26 &1.49 \\
		5 &1.07 &0.93                 &0.89 &1.10 &0.93                 &0.79                 &0.98                 &0.90 &1.10 \\
		6 &0.79 &0.67                 &0.65 &0.77 &0.66                 &0.67                 &0.83                 &0.69 &0.78 \\
		7 &0.55 &0.48                 &0.43 &0.58 &0.43                 &0.78                 &0.92                 &0.50\tnote{h} &0.58 \\
		8 &0.41 &0.33                 &0.28 &0.40 &0.31                 &0.69                 &\multicolumn{1}{c}{-}&0.46\tnote{h} &0.42\tnote{i} \\
		9 &0.32 &0.24                 &0.22 &0.30 &\multicolumn{1}{c}{-}&\multicolumn{1}{c}{-}&\multicolumn{1}{c}{-}&\multicolumn{1}{c}{-}&0.33\tnote{i} \\
		10&0.26 &\multicolumn{1}{c}{-}&0.22 &0.20 &\multicolumn{1}{c}{-}&\multicolumn{1}{c}{-}&\multicolumn{1}{c}{-}&\multicolumn{1}{c}{-}&0.19\tnote{j} \\ \bottomrule
	\end{tabular}
	\begin{tablenotes}
		\item[a] Ref \citenum{DSRG2025}, DMRG-based driven similarity renormalization group third-order perturbation theory, def-TZVP basis set, DMRG $M=600$, UB3LYP/6-31G(d,p) geometry\cite{PolyAceneGASGeom}.
		\item[b] Ref \citenum{ecmrciDMRG}, DMRG-based external contracted multireference configuration interaction with singles and doubles (ec-MRCISD) and Pople correction (Q), ANO-L-VTZP(C)/ANO-S-MB(H) basis set, DMRG $M=1000$,  UB3LYP/6-31G(d,p) geometry\cite{PolyAceneGASGeom}.
		\item[c] Ref \citenum{PolyAceneGASGeom}, generalized active space-based multiconfigurational pair-density functional theory (PDFT) with tPBE on-top functional, 6-31+G(d,p) basis set, WFP-3 GAS partition, UB3LYP/6-31G(d,p) geometry\cite{PolyAceneGASGeom}.
		\item[d] Ref \citenum{PolyAceneGASGeomCCSDT}, focal point analysis, B3LYP/cc-pVTZ geometries.
		\item[e] Ref \citenum{ICEFRNEPT2}, iterative configuration expansion (ICE) based full rank (FR) NEVPT2, cc-pVTZ basis set, UB3LYP/6-31G(d) geometry\cite{ACI2018}.
		\item[f] Ref \citenum{ASCIMRPT2}, cc-pVTZ basis set, ASCISCF(2)/cc-pVDZ geometry\cite{ASCISCF_OptGeom}.
		\item[g] cc-pVTZ basis set, geometries of naphthalene ($k=2$) and k-acene ($k=3,4,\cdots,10$) are taken from Refs \citenum{ACI2018} and \citenum{ASCISCF_OptGeom} respectively.
		\item[h] Extrapolated value with $C_{\text{min}}=\{5.0,3.0,1.5,0.9\}\times 10^{-5}$.
		\item[i] Extrapolated value with $C_{\text{min}}=\{5.0,3.0,1.5\}\times 10^{-5}$.
		\item[j] Extrapolated value with $C_{\text{min}}=\{7.0,5.0,3.0\}\times 10^{-5}$.  
	\end{tablenotes}
\end{threeparttable}
\label{table:polyacene}
\end{sidewaystable}

\clearpage
\newpage

\subsection{\ce{Fe} Porphyrin}
\begin{figure}[!ht]
	\includegraphics[width=0.9\textwidth]{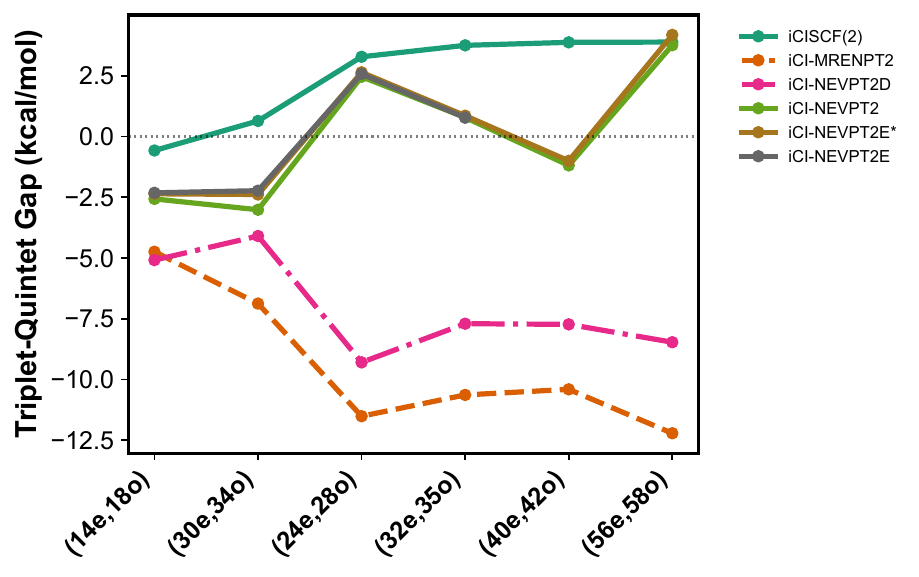}
	\caption{Triplet-quintet gaps ($\Delta E=E_Q-E_T$ in kcal/mol) by iCISCF(2) and iCI-MRPT2 with various active spaces and cc-pVDZ basis set.
	}\label{fig:FeP_iCIMRMP2}
\end{figure}

The low-lying spin states of Fe(II) porphyrin (FeP) complexes are difficult to describe, for the results
are rather sensitive to the choice of active spaces. Facilitated by the iCAS (imposed complete active space) approach\cite{iCAS},
the following six active spaces were tested before\cite{iCISCF}:
\begin{itemize}
	\item (14e,18o) : Fe 3d4s4p4d + 4 sp$^2$-PHO (projected hybrid orbital\cite{PHO}) of \ce{N}.
	\item (30e,34o) : Fe 3d4s4p4d + 12 sp$^2$-PHO of \ce{N} + 8 sp$^2$-PHO of \ce{C}
	\item (24e,28o) : Fe 3d4s4p4d + 4 sp$^2$-PHO  of \ce{N} + 10 porphyrin $\pi$ orbitals.
	\item (32e,35o) : Fe 3d4s4p4d + 4 sp$^2$-PHO  of \ce{N} + 17 porphyrin $\pi$ orbitals.
	\item (40e,42o) : Fe 3d4s4p4d + 4 sp$^2$-PHO  of \ce{N} + 24 porphyrin $\pi$ orbitals.
	\item (56e,58o) : Fe 3d4s4p4d + 12 sp$^2$-PHO  of \ce{N} + 8 sp$^2$-PHO  of \ce{C} + 24 porphyrin $\pi$ orbitals.
\end{itemize}
The iCISCF orbitals with $C_{\text{min}}=5\times 10^{-5}$
 were directly taken from Ref.~[\citenum{iCISCF}]. However, the inner-space ENPT2 calculations 
were redone with smaller $C_{\text{min}}$ (i.e. $\{5.0,3.0,1.5,0.9,0.7\}\times 10^{-5}$; see Tables S9 and S10
in the Supporting Information), so as to make the iCISCF(2) energies even closer to the CASSCF ones. 
As shown in Fig.~\ref{fig:FeP_iCIMRMP2}, iCISCF(2)/CASSCF with CAS(14e,18o) fails to produce the correct energy ordering of the triplet and quintet states. 
Going from CAS(24e,28o) to CAS(56e,58o), the iCISCF(2) triplet--quintet gap $\Delta_{QT}$ converges to 3.9~kcal/mol.
In more detail, the $\Delta_{QT}$ values with CAS(40e,42o) and CAS(56e,58o) [which has additional 
16 \ce{N}--\ce{C} $\sigma$ and $\sigma^\ast$ orbitals relative to CAS(40e,42o)]
 are essentially the same, whereas those 
with CAS(30e,34o) and CAS(56e,58o) [which has additional 
additional 24 porphyrin $\pi$ orbitals relative to CAS(30e,34o)]
differ by 3.3 kcal/mol, 
indicating that the additional 14 porphyrin $\pi$ orbitals (denoted as $\pi_{14}$) when going from CAS(30e,34o) to CAS(40e,42o)
 are more important than the 16 \ce{N}--\ce{C} $\sigma$ and $\sigma^\ast$ orbitals (denoted as $\sigma_{16}$) 
 when going from CAS(14e,18o) to CAS(30e,34o) at the iCISCF(2)/CASSCF level. 

However, when dynamic correlation is taken into account, the $\Delta_{QT}$ values vary strongly with respect to the active spaces
and oscillatie around zero. 
This is related to the fact that high-spin states usually have less dynamic correlation than low-spin states. 
On the other hand, treating the $\sigma_{16}$ orbitals  variationally (in the 
active space) [or perturbatively (in the outer space)], but  
the $\pi_{14}$ orbitals perturbatively (or variationally) is unbalanced. Indeed,  
when the two sets orbitals are treated in the same way, either both perturbatively (CAS(24e,28o)) or both variationally (CAS(56e,58o)),
consistent $\Delta_{QT}$ values can be obtained. 

Both iCI-MRENPT2 and iCI-NEVPT2D predict overly negative $\Delta_{QT}$ values, 
reaffirming that they are not suited for dynamic correlation of the external space. 

Due to the use of different basis sets, active spaces, and treatments of dynamic correlation,
the available theoretical results for the $\Delta_{QT}$ values of FeP span a wide range, see
Table \ref{table:FeP}. There is no clear hint on the basis set effects
by comparing the two Stochastic-CASSCF+TDCSD\cite{StochasticGASSCF_FeP} and the two ICE-FR-NEVPT2\cite{ICEFRNEPT2} calculations.

\begin{table}[htbp]
	\small
	\centering
	\caption{Comparison of the Triplet-Quintet Gap (in kcal/mol) of \ce{Fe} Porphin from Various Theoretical Methods}
\begin{threeparttable}
	\begin{tabular}{lccS} \toprule
 \multicolumn{1}{c}{method}
	& basis set
	& active space
	& \multicolumn{1}{c}{$\Delta E$}  \\ \toprule
	DMRG-PDFT:tPBE\tnote{a}
	& ANO-RCC\tnote{b}
	& (34e,35o)\tnote{c}
	& 9.7  \\
	DMRG-PDFT:ftPBE\tnote{a}
	& ANO-RCC\tnote{b}
	& (34e,35o)\tnote{c}
	& 16.1  \\
	Stochastic-CASSCF+RASCI\tnote{d}
	& ANO-RCC\tnote{e}
	& (32e,34o)\tnote{f}
	& 7.0(1) \\
    Stochastic-CASSCF+Stochastic SplitGAS\tnote{g}
	& ANO-RCC\tnote{e}
	& (32e,34o)\tnote{f}
	& 8.0(1) \\
	Stochastic-CASSCF+TDCSD\tnote{h}
	& ANO-RCC\tnote{e}
	& (32e,34o)\tnote{f}
	& 2.6 \\
	Stochastic-CASSCF+TDCSD+F12\tnote{h}
	& ANO-RCC\tnote{e}
	& (32e,34o)\tnote{f}
	& 5.8 \\
	FCIQMC-CASPT2\tnote{i}
	& ANO-RCC\tnote{e}
	& (26e,27o)\tnote{j}
	& -2.2 \\
	ICE-FR-NEVPT2\tnote{k}
	& cc-pVTZ
	& (30e,29o)\tnote{l}
	& -4.94 \\
	ICE-FR-NEVPT2\tnote{k}
	& cc-pVQZ
	& (30e,29o)\tnote{l}
	& -7.69 \\ \midrule
	iCI-NEVPT2\tnote{m}
	& cc-pVDZ
	& (24e,28o)
	& \multicolumn{1}{l}{2.46(1)/2.64(1)/2.57(1)}\\
	& cc-pVDZ
	& (32e,35o)
	& \multicolumn{1}{l}{0.78(9)/0.86(9)/0.77(9)}\\
	& cc-pVDZ
	& (40e,42o)
	& \multicolumn{1}{l}{-1.2(4) /-1.0(4) /-}\\
	& cc-pVDZ
	& (56e,58o)
	& \multicolumn{1}{l}{3.8(9) /4.2(9)  /-}\\ \bottomrule
	\end{tabular}
\begin{tablenotes}
	\item[a] Ref \citenum{FePPDFT}. DMRG-based PDFT. $D_{\text{4h}}$ geometry.
	\item[b] ANO-RCC basis set\cite{ANO1,ANO2,ANO3} contracted to [7s6p5d3f2g1h] for \ce{Fe}, to [4s3p2d1f] for \ce{C} and \ce{N}, and to [3s1p] for \ce{H}.
	\item[c] Fe valence and double d-shell orbitals, one \ce{Fe}-\ce{N} $\sigma$-bonding orbital, and 24 porphyrin $\pi$ orbitals.
	\item[d] Ref \citenum{StochasticGASSCF_FeP}. 
	Correlating 96 electrons into 159 orbitals.  
	$D_{\text{4h}}$ geometry.
	\item[e] ANO-RCC-VTZP basis set\cite{ANO1,ANO2,ANO3}.
	\item[f] Fe valence and double-shell orbitals, all porphyrin $\pi$ orbitals and the $\sigma$ donating orbitals.
	\item[g] Ref \citenum{SplitGAS_FeP}. 96 correlated electrons in 159 orbitals.  $D_{\text{4h}}$ geometry.
	\item[h] Ref \citenum{StochasticGASSCF_FeP}. TDCSD denotes the tailored distinguishable cluster method with singles and doubles.  $D_{\text{4h}}$ geometry.
	\item[i] Ref \citenum{FCIQMC_CASPT2}.  $D_{\text{4h}}$ geometry.
	\item[j] Active space pruned from CAS(32e,34o) based on natural occupation numbers.
	\item[k] Ref \citenum{ICEFRNEPT2}. $D_{\text{4h}}$ geometry.
	\item[l] Fe valence and double d shell orbitals, Fe 3s3p4s4p orbitals, one \ce{Fe}-\ce{N} $\sigma$-bonding orbital, 
four Gouterman orbitals, two $\sigma$ ligand orbitals, and four porphyrin $\pi$ orbitals.
	\item[m] iCI-NEVPT2/iCI-NEVPT2E*/iCI-NEVPT2E. In parentheses are extrapolation uncertainties. $D_{\text{2h}}$ geometry.
\end{tablenotes}
\end{threeparttable} \label{table:FeP}
\end{table}

\clearpage
\newpage

\section{CONCLUSIONS and OUTLOOK}\label{sec:conclusion}
Based on the fact that selected iCI can construct very efficiently a compact but high-quality variational space for static correlation,
a family of MRPT2s has been developed to account for the residual dynamic correction in strongly correlated systems of electrons. While they differ
in the choice of zeroth-order Hamiltonians, a unified implementation of them has been achieved, by making use of the infrastructure
in the MetaWave platform. Additional unique features can be summarized as follows: 
\begin{enumerate}[(1)]
\item All variants of iCI-MRPT2 are fully spin adapted.
\item The use of partial contraction avoids completely higher than second-order RDMs.
\item The algorithmic optimizations for the three uncontracted subspaces $Q_{,ij}^{(2)}$, $Q_{a,j}^{(2)}$
and $Q_{ab}^{(2)}$ [which involve at most two active orbitals, thereby permitting in-core storage of the active orbital configurations (i.e.,
second-order residues)] are nearly ten times faster than a standard implementation.
\item The LCUT approximation to the first-order amplitudes of iCI-MRMP2 and iCI-NEVPT2E is sufficiently accurate
for practical applications.
\item Sequential outer-space (via $Q_{\mathrm{min}}$) and inner-space (via $C_{\mathrm{min}}$) extrapolations
can be made to approach the CAS-based counterparts.
\item iCI-NEVPT2 is strictly size-consistent, whereas iCI-NEVPT2E* is numerically size-consistent. 
\end{enumerate} 
These features render iCI-MRPT2 applicable to active spaces around (60e,60o) and full orbital spaces up to (300e,1600o).
Nevertheless, the following points need to be resolved in the near future: (a) MPI parallelization;
(b) extension to multi-states; (c) systematic examination of the extended Dyall Hamiltonian;
(d) combination with relativistic Hamiltonians. Work along these directions is being carried out at our laboratory. 

\section*{ACKNOWLEDGMENTS}
This work was supported by National Natural Science Foundation of China  (Grant Nos. 22503051 and 22373057),  National Postdoctoral Science Foundation of China (Grant Nos. 2026M791090 and 2026T190280) and Shandong Postdoctoral Science Foundation and by Qingdao Postdoctoral Applied Research Project. 

\section*{Supporting Information}
Extrapolated iCI-MRPT2 energies of \ce{Cr2}, polyacenes and \ce{Fe} porphyrin.
For \ce{Cr2}, the total energies  are collected in Tables S1 and S2, and the spectroscopic constants derived from the Morse fit are given in Table S3.
For polyacenes, the total energies and perturbation corrections are collected in Tables S4--S6, the adiabatic singlet--triplet gaps in Tables S7--S9, and the sizes of the primary spaces for the singlet and triplet states in Tables S10 and S11.
For \ce{Fe} porphyrin, the sizes of the primary spaces 
are given in Tables S12 and S13, and the total energies, perturbation corrections and triplet--quintet gaps are collected in Tables S14--S16.

\section*{Data and Software Availability}
The original data, testing and analyzing scripts, typical input and output file have been made publicly available on the GitHub repository SI\_MRPT2\cite{SIMRPT2}.
MetaWave, distributed as part of the BDF software, is closed-source.
Access to the MetaWave package may be granted upon request to the authors.

\section*{Conflicts of interest}
There are no conflicts to declare.



\newpage

\bibliography{iCI}

\end{document}